\documentclass[sigconf]{acmart}
\usepackage{enumitem} 
\usepackage{tikz}
\usetikzlibrary{calc}
\usepackage{graphicx}
\usepackage{subcaption}
\usepackage{makecell,multirow}
\usepackage{float}
\usepackage{stfloats}

\AtBeginDocument{%
  }

\copyrightyear{2026}
\acmYear{2026}
\setcopyright{cc}
\setcctype{by}
\acmConference[DIS '26]{Designing Interactive Systems Conference}{June 13--17, 2026}{Singapore, Singapore}
\acmBooktitle{Designing Interactive Systems Conference (DIS '26), June 13--17, 2026, Singapore, Singapore}
\acmDOI{10.1145/3800645.3813003}
\acmISBN{979-8-4007-2563-0/2026/06}

\usepackage{xstring}

\usepackage{fvextra}
\usepackage{xcolor}
\usepackage{framed}
\definecolor{lightgray}{RGB}{245,245,245}
\colorlet{shadecolor}{lightgray}

\newenvironment{codebox}{%
\par\noindent
  \MakeFramed{\advance\hsize-\width \FrameRestore}%
}{%
  \endMakeFramed
}

\DefineVerbatimEnvironment{myverbatim}{Verbatim}{
  breaklines=true,   
  breakanywhere=true, 
  fontsize=\small     
}

\usepackage{xcolor}

\definecolor{lightgray}{RGB}{245, 245, 245}
\definecolor{lightyellow}{RGB}{255, 255, 230}
\definecolor{lightgreen}{RGB}{230, 255, 230}
\definecolor{lightblue}{RGB}{230, 240, 255}
\newcommand{\condition}[1]{%
  \IfStrEqCase{#1}{%
    {none}{\colorbox{lightgray}{\textsc{no ai}}}%
    {plan}{\colorbox{lightgreen}{\textsc{ai plan}}}%
    {revision}{\colorbox{lightblue}{\textsc{ai revision}}}%
    {draft}{\colorbox{lightyellow}{\textsc{ai draft}}}%
  }[\textsc{unknown}]
}

\begin{document}

\title{From Planning to Revision: How AI Writing Support at Different Stages Alters Ownership}

\author{Katy Ilonka Gero}
\email{katy.gero@sydney.edu.au}
\orcid{1234-5678-9012}
\affiliation{%
  \institution{University of Sydney}
  \city{Sydney}
  \state{NSW}
  \country{Australia}
}

\author{Tao Long}
\email{long@cs.columbia.edu}
\affiliation{%
  \institution{Columbia University}
  \city{New York City}
  \state{NY}
  \country{USA}
}

\author{Carly Schnitzler}
\email{cschnit1@jh.edu}
\affiliation{%
  \institution{Johns Hopkins University}
  \city{Baltimore}
  \city{MD}
  \country{USA}}

\author{Paramveer S. Dhillon}
\email{dhillonp@umich.edu}
\affiliation{%
  \institution{University of Michigan}
  \city{Ann Arbor}
  \state{MI}
  \country{USA}
}


\begin{abstract}
  Although AI assistance can improve writing quality, it can also decrease feelings of ownership. Ownership in writing has important implications for attribution, rights, norms, and cognitive engagement, and designers of AI support systems may want to consider how system features may impact ownership. We investigate how the stage at which AI support for writing is provided (planning, drafting, or revising) changes ownership. In a study of short essay writing (between subjects, n = 253) we find that while any AI assistance decreased ownership, planning support only minimally decreased ownership, while drafting support saw the largest decrease. This variation maps onto the amount of text and ideas contributed by AI, where more text and ideas from AI decreased ownership. Notably, an AI-generated draft based on participants' own outline resulted in significantly more AI-contributed ideas than AI support for planning. At the same time, more AI contributions improved essay quality. We propose that writers, educators, and designers consider writing stage when introducing AI assistance.
\end{abstract}




\maketitle

\section{Introduction}
Artificial intelligence (AI) systems such as large language models (LLMs) are increasingly being adopted to support writing tasks \cite{liang2025WidespreadAdoptionLarge}, but the long-term implications of this are only just starting to be understood. Numerous studies have found that AI writing support can have a positive impact on writing quality,  efficiency, and confidence \cite{li2024ValueBenefitsConcerns, noy2023ExperimentalEvidenceProductivity,dhillon2024ShapingHumanAICollaboration}, but there is also evidence that AI writing support decreases feelings of ownership \cite{dhillon2024ShapingHumanAICollaboration, lee2022CoAuthorDesigningHumanAI, kobiella2024IfMachineGood,long2025longitudinal,li2024ValueBenefitsConcerns}, decreases population-level diversity \cite{doshi2024GenerativeAIEnhancesa, anderson2024HomogenizationEffectsLarge, li2024ValueBenefitsConcerns}, decreases cognitive engagement \cite{kosmynaYourBrainChatGPT, li2024ValueBenefitsConcerns} and may result in deskilling \cite{varanasi2025AIRivalryCraft}. 
In this work, we focus on how AI writing support impacts feelings of ownership.
As AI technologies have become more capable, concerns about their impact on ownership have grown.

Designers and researchers are creating new interaction paradigms for AI-assisted writing support \cite{masson2024DirectGPTDirectManipulation, masson2025TextoshopInteractionsInspired}, and so they are also interested in how to design for ownership. For instance, the HaLLMark system \cite{hoque2024HaLLMarkEffectSupporting} visualizes a writer's interaction with an LLM  while DraftMarks \cite{siddiqui2025DraftMarksEnhancingTransparency} uses process traces to support readers in interpreting the provenance of AI-assisted texts. As new interfaces for human-AI co-creation more generally are being developed at a rapid pace~\cite{hu2025DesigningInteractionsGenerative}, understanding what impacts ownership will allow designers to make informed decisions.
While there is good evidence that AI writing support can decrease feelings of ownership, precisely why or how is less understood. Studies have found a range of factors can influence ownership, including amount of AI-generated text \cite{lee2022CoAuthorDesigningHumanAI, dhillon2024ShapingHumanAICollaboration} and amount of AI-contributed ideas \cite{kobiella2024IfMachineGood, draxler2024AIGhostwriterEffect}.

Drawing on insights from writing studies \cite{sommers1980RevisionStrategiesStudent, flower1981CognitiveProcessTheory}, we bring nuance to this question by investigating how AI support at \textbf{different writing stages} affects ownership. We separate out a writing task into planning, drafting, and revision, a common and widespread practice among both writing researchers \cite{flower1981CognitiveProcessTheory, johnhayes2000NewFrameworkUnderstanding} as well as writing instructors \cite{writingprocessUCBerkeley, writingprocessMITCMS, writingprocessKU, writingprocessEBSCO}.
We isolate AI support in each of these stages to investigate if feelings of ownership are most impacted by AI support at a particular stage, offering empirical grounding for designers navigating tradeoffs between assistance and agency.
We have three main research questions:

\smallskip

\noindent \textit{How does the stage at which AI support is offered:}

\begin{itemize}
    \item RQ1: \textit{affect a writer's feelings of ownership?}
    \item RQ2: \textit{change how a writer attributes text and ideas to themselves versus the AI?}
    \item RQ3: \textit{affect the final essay's quality?}
\end{itemize}

To investigate these questions, we run an online, between-subjects experiment (n=253) in which participants are asked to write a 300-word argumentative essay in three stages. Participants are first asked to produce an outline, then a draft, and then spend time revising. Our experiment has four conditions: no AI support (\condition{none}), AI support for planning (\condition{plan}), AI-generated draft based on participant outline (\condition{draft}), and AI support for revision (\condition{revision}). We designed our conditions to match how people typically use AI support for writing in each stage. 
We measure if and how participants actually make use of the provided AI support. We evaluate 
participants' feelings of ownership (as well as related concepts of agency, intent, and effort), attribution of text and ideas, and essay quality.

We find that ownership is highest with no AI support, lower for planning support, lower again for revision support, and lowest for AI-generated draft. Participants' ownership feelings map cleanly on their perception of AI attribution, where higher attribution of text and ideas to the AI results in decreased feelings of ownership. Interestingly, text and idea attribution appear to be coupled in our experiment across all conditions.  

Notably, even though the AI-generated draft is based on participants' own outline, and participants edited the AI draft more than in the other conditions, this still resulted in the highest amount of ideas attributed to the AI (27\% for \condition{draft}, compared to 16\% for \condition{revision} and 11\% for \condition{plan}), contrary to our expectations that the AI draft would contribute much text but fewer ideas. 
We also find that even when participants request AI ideas during outlining, they often find these ideas redundant with their own prior ideas or choose to rely on their own ideas instead; this is in contrast to AI suggestions for revisions which are incorporated much more often. This suggests that AI support for the planning stage may have the least impact, both on writing content and writer ownership, and has implications for designers, educators, and end users incorporating AI into any creative process.

We find that essay quality is negatively correlated with ownership, with the highest essay quality for \condition{draft} and lowest for \condition {none}. This suggests an ownership-quality tradeoff, in which an AI-generated draft buys high quality essays at a noticeable ownership cost. Although this is in line with previous studies \cite{dhillon2024ShapingHumanAICollaboration}, we consider how the structure of the task and our grading rubric may be contributing to this finding.

In our discussion, we  consider how our results provide evidence that writing is a "mode of learning" \cite{emig1977WritingModeLearning} and what might explain the differences in ownership results between planning and revision support. Based on our findings we present three design implications: temporally fading out AI interactions, generating incomplete text that requires rewriting, and making the chatbox editable. Overall, this work provides empirical evidence for how writing support at different stages has different impacts on ownership and attribution, and we provide suggestions for future work in the design of AI support tools.





\section{Related Work}

We start by drawing on the long intellectual history of ownership in philosophy and then psychology, finally attending to ownership as it relates to writing specifically. Then we consider the more recent literature on AI's impact on ownership and how researchers and designers are redesigning AI support for writing. 
Finally, we give a brief overview on the literature on writing stages.

\subsection{Psychological Feeling of Ownership}

The issues of property and ownership have been considered by philosophers for centuries \cite{sep-property}. A useful entry point is to consider the materialist versus idealist theories, which demonstrate variation in how feelings of ownership can develop. Central to the materialist view is Locke's labor theory, which states that 
when one mixes their labor with natural resources, the resulting object becomes their property \cite{Locke1689}. This \textit{material} theory of ownership focuses on physical, material states as determinants of reality. On the other hand, Hegel's \textit{idealist} theory considers how one's "will" can be embodied in objects, aligning ownership more with the investment of one's identity and values \cite{Hegel2015-HEGTPO-28}. 

Writing studies scholarship stemming from the cognitive turn in the 1970s and 80s concretizes these ideas. Because writing forces a slowing of thought, it encourages a “shuttling among past, present, and future,” enabling meaning-making through analysis and synthesis \cite{emig1977WritingModeLearning}. This “shuttling” between our three tenses of experience is a special temporal zone that writing occupies in cognition, and is what allows writers to set their own goals within a particular writing task \cite{emig1977WritingModeLearning, flower1981CognitiveProcessTheory}. The recursive and intermediate goal setting that writers do contributes heavily to feelings of ownership—what Flower and Hayes (1981) described as using the process of writing to think through something as well as produce a material product~\cite{flower1981CognitiveProcessTheory}.

Time and labor investments in a process can lead to increased perceived value and psychological ownership, a cognitive bias colloquially known as the “Ikea effect” \cite{norton2012IKEAEffectWhen}. Ownership is a cognitive-affective state; as \citet{pierce2003StatePsychologicalOwnership} find, it is both a state that one is aware of through "intellectual perception" and a state that one emotionally feels. Perceived ownership over something is knowing that one chose to put time and work in (“control” and “investment of self”) and feeling that these things are valuable (“intimate knowing”). 
In the act of writing, according to cognitive models, 
there is agency or “control” in goal-setting and goal-recreation, “investment of self” in the recursiveness of this process, and “intimate knowing” of understanding one has learned something \cite{pierce2003StatePsychologicalOwnership, flower1981CognitiveProcessTheory}. Though, as both psychologists and writing studies scholars note, these routes are entangled: they are "distinct, complementary, and additive" and "it is not clear whether some routes are more effective at generating psychological ownership than others" \cite{pierce2003StatePsychologicalOwnership}. 

These processes toward ownership in writing are all underwritten by time--both for task completion and elision in writing stages--something AI support reduces significantly. In this work, we investigate the various "routes" towards ownership of a piece of writing by comparing the contribution of ideas (idealist) versus text (materialist) from AI support, but understand that the role of time in both task completion (materialist) and writing process (idealist) is significant, hence splitting out writing into distinct temporal stages.

\subsection{AI's Impact on Ownership}

Early work on AI assistance for writing found that users may avoid such assistance due to decreased feelings of ownership over the final artifact \cite{gero2019MetaphoriaAlgorithmicCompanion}. 
From a "materialist" perspective, it has been found that more AI-generated text included in the final artifact decreases ownership scores \cite{lee2022CoAuthorDesigningHumanAI, dhillon2024ShapingHumanAICollaboration}, and
that shorter prompts for the AI (which could be interpreted as less labor) decreased feelings of ownership \cite{joshi2024WritingAILowers}. From an "idealist" perspective, it has been found that less control or agency, especially over "core content", decreases ownership \cite{kobiella2024IfMachineGood, draxler2024AIGhostwriterEffect, reza2025CoWritingAIHumana}, although \citet{yeh2024GhostWriterAugmentingCollaborative} found that increased agency in the co-writing process alone did not dramatically affect ownership. 
Given a clear interest in preserving ownership over writing artifacts, designers have begun considering this when building new writing interfaces. In the next section, we consider how AI writing assistance is being designed, and how those design decisions impact ownership.

\subsection{Designing for AI Writing Assistance}

Early AI writing tools often worked in an autocomplete design paradigm \cite{chen2019GmailSmartCompose, clark2018CreativeWritingMachine}; the rise of large language models shifting the norm towards a chatbot design. As AI abilities increase, researchers have explored new interaction paradigms, such as direct manipulation \cite{masson2024DirectGPTDirectManipulation, masson2025TextoshopInteractionsInspired}, material writing \cite{calderwood2025PhraselettePoetsProcedural}, surfacing many options at once \cite{suh2024LuminateStructuredGeneration, gero2024SupportingSensemakingLarge}, and methods to preserve provenance \cite{hoque2024HaLLMarkEffectSupporting, siddiqui2025DraftMarksEnhancingTransparency}. \citet{buschek2024CollageNewWriting} proposes \textit{collage} as a concept to analyze, construct, and critically evaluate new user interfaces for AI writing tools. Collage makes sense of the trend towards fragmented text made possible by sophisticated AI-generated text as users may shift from manual writing to a more editorial role. We see the concept of collage as highlighting the way in which AI writing tools embrace the ability to generate or change text rapidly, which can lead to writers accepting text without reading it carefully or integrating it into their conception of the writing's purpose or value, and ultimately shape their sense of ownership via both the materialist and idealist routes. 

While some AI writing tools are intended to be general purpose, others focus on specific users or writing tasks. Responding to this, \citet{reza2025CoWritingAIHumana} investigate how ownership perceptions vary across over 100 different AI writing tool designs, finding that writers seek differing kinds of AI support depending on if they are focused on the content or form of their writing (similar to the findings in \cite{gero2023SocialDynamicsAI}). However, while the work by \citet{reza2025CoWritingAIHumana} highlights that research has focused on writing support across the stages of writing, it leaves open questions about the actual impact of the writing stage, as studies tend to focus on improving support in just one stage. And despite new designs opening up the ways in which people can use AI for writing, we still know little about how typical usage patterns are impacting ownership. We build on prior work by empirically testing how ownership varies depending on the stage in which AI support is used; our study investigates more typical usage patterns with an eye towards how this might lead us to new designs.

As designers explore this expanding design space, we lack empirical work about how design choices affect core aspects of the writing experience. While we know that AI writing support can decrease ownership, we don't yet understand how specific \textit{design decisions}--such as when support occurs in the writing process--shape this outcome. We address this gap by investigating how AI support at different writing stages impacts ownership, attribution of text and ideas, and writing quality. This provides evidence for designers hoping to embed certain values within their designs.

\subsection{Writing Stages}

Many writing tasks are segmented into planning, drafting, and revising processes, although these processes are not necessarily linear and can also be hierarchical \cite{johnhayes2000NewFrameworkUnderstanding, flower1981CognitiveProcessTheory}.\footnote{For instance, one may plan, draft, and revise an outline, which is itself a larger plan that is used to write a longer draft. In this way, planning, drafting, and revising are writing \textit{processes}, rather than ordered stages.} That said, schools often teach students to structure their writing with these processes \cite{writingprocessUCBerkeley, writingprocessMITCMS, writingprocessKU, writingprocessEBSCO} and these processes, more generally, align well with models of creative or complex processes such as Amabile’s model of creativity (preparation, generation, validation) \cite{Amabile1983TheSP}, the design thinking framework (ideate, prototype, test) \cite{designthinking}, and Zimmerman's model of self-regulated learning (forethought, performance, self-reflection) \cite{zimmerman2002BecomingSelfRegulatedLearner}, lending weight to them as a model applicable across a variety of tasks. Many studies have used the planning-drafting-revision framework to investigate AI writing assistance \cite{reza2025CoWritingAIHumana,lee2024DesignSpaceIntelligent,gero2022SparksInspirationScience,umarova2025HowProblematicWriterAI}, suggesting its utility as an analytical framework.

Writing studies scholarship acknowledges both the practical utility and shortcomings of the ``stage models of writing,'' which we define here as pre-writing or outlining, writing or drafting, and re-writing or revising~\cite{flower1981CognitiveProcessTheory,sommers1980RevisionStrategiesStudent}. These stages remain useful for experiment design, as they mirror when typical pedagogical interventions in writing occur~\cite{sommers1980RevisionStrategiesStudent}. At the same time, these stages are artificial linear distinctions and ultimately elided and recursive in practice—all writing is planning, all writing is revising, all writing (as the saying goes) is rewriting.

\section{Experiment Design}

We ran an online, between-subjects experiment (n=253) in which participants were asked to write a short argumentative essay on a selected topic. Participants were required to first plan their essay by writing an outline, then produce a draft, and then spend time revising their draft. In order to investigate how AI support for different writing stages affects ownership, participants experienced one of four conditions: no AI support (referred to as \condition{none}), AI support for planning only (\condition{plan}), AI-generated drafting based on their own outline written in the planning stage (\condition{draft}), and AI support for revision only (\condition{revision}). Figure~\ref{fig:studyflow} shows the study flow at a high level. The experiment was run on the Prolific platform in August 2025.
This experiment was approved by IRB.

\subsection{Protocol}

\begin{figure*}
\centering
\includegraphics[width=\textwidth]{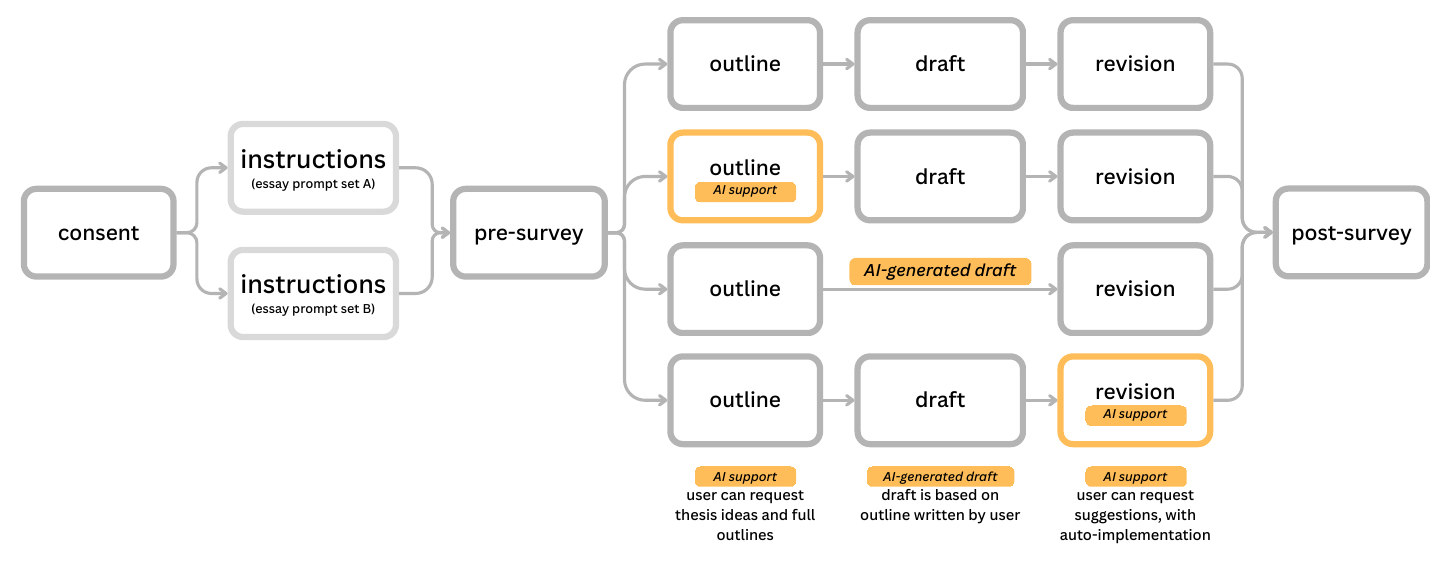}
\caption{Study flow: all participants had to first outline, then draft, then spend time revising their essay. Participants were split into four conditions: no AI support, AI support for outlining (thesis and whole outline ideas), AI-generated draft based on their outline, or revision suggestions (there were three revision tools: argument improver, writing clarity, and proof-reader).}
\label{fig:studyflow}
\Description{Flow chart of the components of the study. From left to right: consent, instructions (essay prompt set A or b), pre-survey, outline, draft, revision, post-survey. For outline, draft, revision, indicates four conditions where AI support is provided for one stage only in each condition; one condition has no AI support.}
\end{figure*}

Participants were first asked to complete a survey that included demographic questions (age, gender, education level, native English speaker), the Need for Cognition scale \cite{cacioppo1982NeedCognition}, technology acceptance, frequency and type of AI usage for writing, efficacy of AI usage for writing, and writing self-efficacy.
Complete survey questions can be found in \autoref{sec:surveyqs}. All questions were either multiple choice or Likert scale questions.

Participants were then instructed that they would be asked to write a 200-300 word argumentative essay on a provided topic. To encourage authentic engagement in the task, participants could pick a topic from a set of three. To ensure a distribution of topics, participants were randomly assigned to one of two sets of topics. Topics were designed to be similar to TOEFL (Test of English as a Foreign Language) and IELTS (International English Language Testing System) prompts, and are listed in \autoref{app:methods} (\autoref{tab:essaytopics}), along with the distribution with which each topic was chosen by participants. Example topics include "\textit{Should public schools ban smartphones during the school day, or permit limited use for learning and emergencies?}" and "\textit{Should parents have the right to genetically edit their unborn children to prevent diseases, or should we ban genetic modifications to preserve natural human diversity?}"

After selecting a topic, participants were moved through three stages of writing: planning (max 10 minutes, min 60 words), drafting (max 15 minutes, min 300 words), and revision (max 10 minutes, min 250 words). Participants had to hit the minimum word count in order to move onto the next stage, unless they hit the maximum time allowed first, in which case they were automatically moved to the next stage.\footnote{Note that the minimum word count for the revision stage was less than that in the draft stage; this was set such that participants could edit down their draft and still submit their essays. They were not required to type an \textit{additional} 250 words.} While participants were told they must revise their draft, we did not actually prevent them from submitting an unchanged draft, as some participants may do their editing during the drafting stage and find it unnatural to be required to change it more. That said, this did mean participants could "skip" the revision stage if they wanted to finish the task quickly. We address this possibility when we analyze the data.

After submitting their revised essay, participants completed another survey that included Likert scale questions about ownership, control, effort, intent, and authorship, as well as sliders (0-100\%) for the percentage of ideas and final text they would attribute to themselves versus the AI. Finally, there were three optional open-ended questions about ownership, attribution, and anything else they would like to share. Complete survey questions can be found in \autoref{sec:surveyqs}.
\autoref{fig:studyflow} shows how participants were moved through the study. Screenshots and complete instructions for each stage can be found in \autoref{sec:screenshots}.

\subsection{Interface Design}

We based our design on a) analysis of large-scale AI-supported writing session datasets \cite{mysore2025PrototypicalHumanAICollaboration}, b) surveys on what people use writing assistants for \cite{laban2024ChatExecutableVerifiable}, c) commercial writing assistant designs,\footnote{Grammarly and Wordtune} and d) the experiences of one author as a writing instructor at a large university. We designed our conditions to mimic typical usage where possible (acknowledging that it can vary widely) while also isolating support in the different stages. We opted for a controlled study, providing specific support tools, rather than unfettered access to a chatbot, in which case participant usage may have been incomparable.\footnote{We suspect that this design, which suggests more of a ``tool'' metaphor, would result in less ownership decrease than the "agent" metaphor suggested by a chatbot, as participants may want to prescribe ownership to an agent than a tool. However, our experiment is not designed to answer such a question, and primarily we intended to focus on ownership changes based on the stage of writing support, not the interface design decisions.}

During the writing task, participants are shown the stage they are in, instructions for that stage, a large textbox in the main area, and a gray panel on the right; see \autoref{fig:plan-condition} for a screenshot of the planning stage. When participants have AI support, the support is provided in the gray panel on the right. 

\subsection{Conditions and AI support}

\begin{figure*}
\centering
\includegraphics[width=\textwidth]{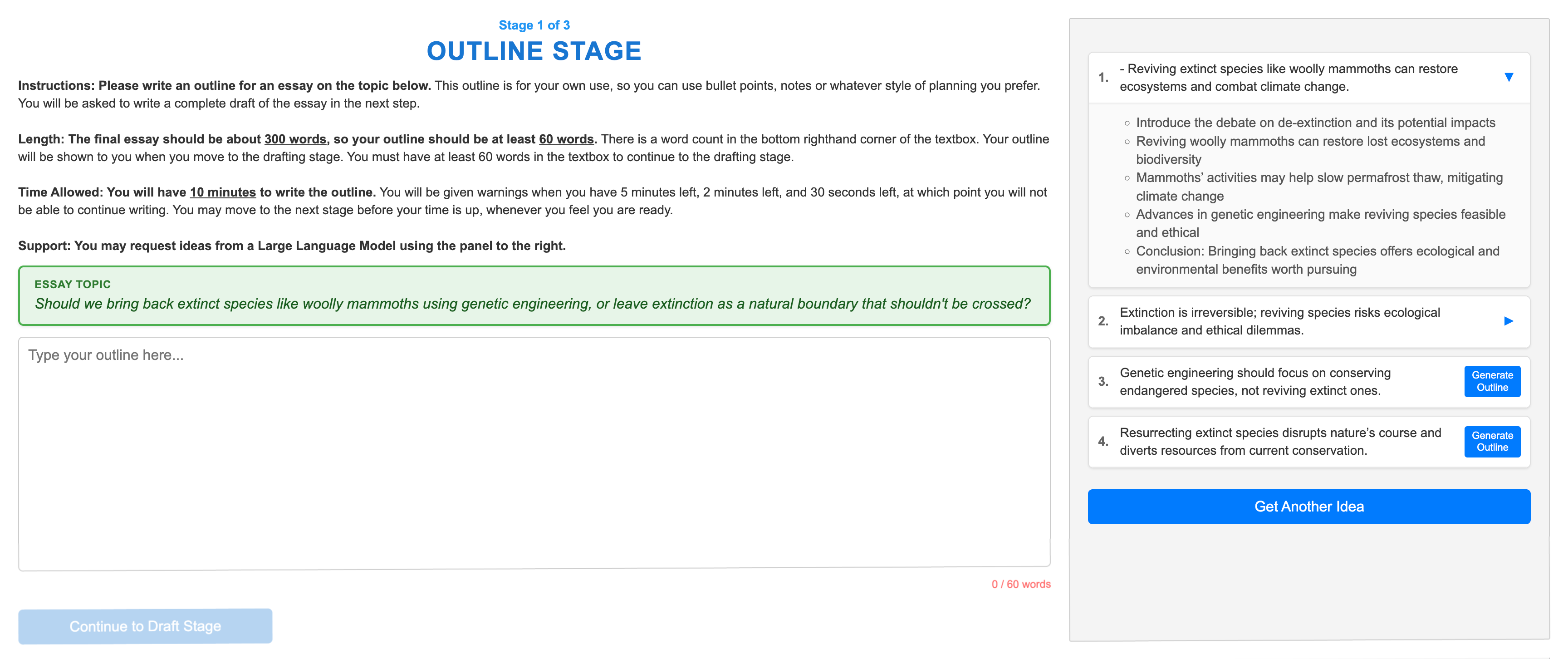}
\caption{A screenshot from the planning condition. Participants wrote their outline in a textbox on the left. A panel on the right allowed participants to request ideas, formulated as single-sentence thesis statements, and then request full outlines based on each idea. The panel starts empty, with just the `Get Idea' button; participants can request as many ideas as they like.}
\label{fig:plan-condition}
\Description{Screenshot is split into two panels. Left panel takes up 70 percent of the page; title is "OUTLINE STAGE", below are instructions for task, essay topic, and then a textbox for writing; word count is available below the textbox. Right panel has four thesis ideas and a "Generate Another Idea" button below; first idea has a fully outline of five points below. Other ideas have a "generate outline" button to their right.}
\end{figure*}

\begin{figure}
    \centering
    \includegraphics[width=0.3\textwidth]{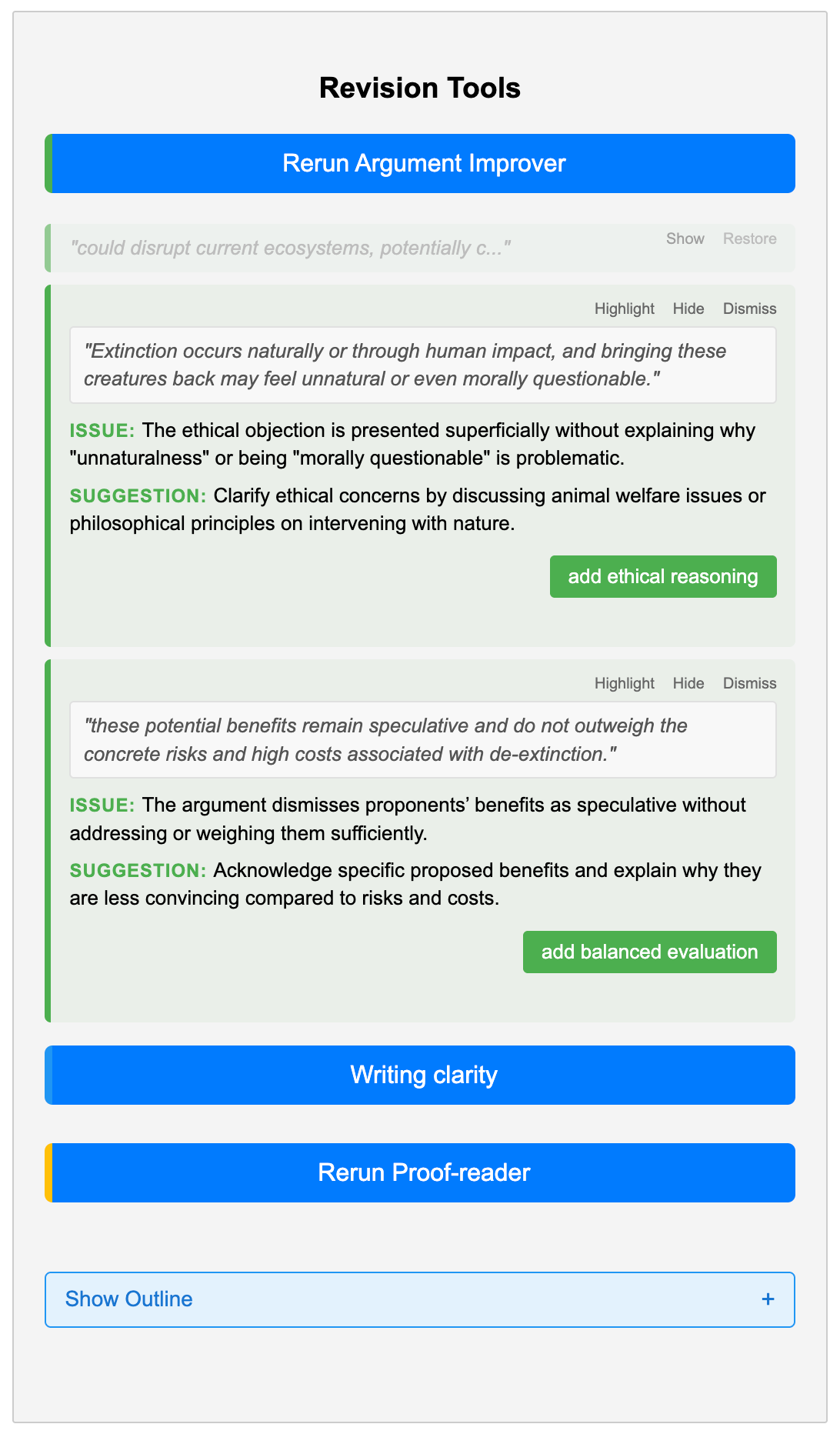}
    \caption{Screenshot of the revision tool panel, showing some suggestions from the "argument improver" tool. There are three revision tools (the three blue buttons) that participants can call; each will generate 2-3 suggestions. Each suggestion includes a quote from the draft, a description of the issue, and a suggested fix. In addition, participants can ask the tool to implement the suggestion for them, which will update the draft in the textbox. All tools can be rerun on the current state of the draft. Finally, participants can see their original outline by clicking the "show outline" button.}
    \label{fig:revision-condition}
    \Description{Skinny panel. Title is "Revision Tools." There are three buttons: "Rerun Argument Improver", "Writing clarity", and "Rerun Proof-reader". Under "Rerun Argument Improver" are two boxes; each has a quote from the text, an issue, and a suggestion. Below all buttons is a "show outline" button.}
\end{figure}

We designed our conditions to match typical AI support for writing in each stage. Participants were randomly assigned to one of four conditions:

\begin{enumerate}
    \item \condition{none}. No AI support was provided. Participants completed all three stages on their own.
    \item \condition{plan}. In the planning stage, participants could request ideas from an AI. When ideas are first requested, three thesis statements are returned. Participants could then request additional thesis statements one at a time. Participants could also request an outline for a generated thesis statement. They could copy and paste ideas from the AI, or write their own. They then had to draft and revise the essay on their own. See \autoref{fig:plan-condition} for a screenshot of this condition.
    \item \condition{draft}. Participants first wrote their own outline during the planning stage, and then clicked a button to get an AI-generated draft based on that outline. They then were given time to revise the draft.\footnote{Supporting the drafting stage in a way comparable to supporting planning and revision was difficult; requesting a full draft based on notes was decided to be the most common drafting-with-LLM technique based on the literature and the author team's experience in writing classrooms. Providing additional options, such as letting participants prompt rewrites of the draft, would shift writers into more of a revision stage than we felt was appropriate. However, we acknowledge the asymmetry between iterative tools for planning and revision, and the one-shot AI draft provided, and suggest future work that could address this concern.}
    \item \condition{revision}. Participants planned and wrote the draft essay on their own. In the revision stage, there were three revision tools: argument improver, writing clarity, and proof-reader. Each tool would generate 2-4 suggestions of how to improve the essay, and had an option to implement the suggested changes. Each tool could be re-run as many times as the participant liked. See \autoref{fig:revision-condition} for a screenshot of the revision tool panel..
\end{enumerate}

Participants were not required to use AI support if it was provided.\footnote{In the case of the AI draft, participants could delete all generated text if they liked.} Accordingly, our primary analyses compare participants by assigned condition regardless of actual engagement with the AI tool. We report AI engagement rates in Table~\ref{tab:engagement_by_stage} and repeat key analyses among active tool users in Appendix~\ref{app:robustness-engagement}. Additional screenshots of the interface can be found in \autoref{sec:screenshots}.

Participants were instructed to do all their typing in the provided text box, and not to use external AI support. We implemented a variety of methods to ensure compliance and data quality, detailed below (\ref{sec:participant-recruitment}). AI support was provided using OpenAI's gpt-4.1-mini model, accessed in August 2025. Full prompts for the model are available in \autoref{app:llmprompts}.

\subsection{Participant Recruitment}
\label{sec:participant-recruitment}
We recruited participants on Prolific and routed them to our web application via Prolific’s URL parameters to record Prolific IDs. We labeled our study as a writing task and restricted access to desktop users and English-speaking adults aged 22–65 residing in the United States, United Kingdom, or Australia. To improve data quality we limited the sample to workers with a high approval history on Prolific (100\%) and at least 100 prior Prolific submissions, and we excluded anyone who had participated in our pilot study.  We followed Prolific’s fair-pay guidance for our compensation. We provided a base payment of \$10 for an estimated 40-minute session ($\sim$\$15/hr). To encourage careful, good-faith writing effort (not just speed), we provided an additional \$5 bonus for the top 5\% of essays based on our independent quality scoring procedure.

We implemented extensive in-browser telemetry (e.g., interaction logs and content provenance checks) and applied pre-specified exclusion criteria to protect the study manipulations. In particular, submissions were excluded if participants used external AI tools outside of their assigned study condition or copy-pasted substantial text from outside the study interface. Further details can be found in \autoref{app:qualitycontrol}. We ended up rejecting 18\% of submissions before our statistical analysis. 
Our final data sample consisted of 253 participants distributed almost evenly across the four study conditions (\condition{none}: n=60; \condition{plan}: n=66; \condition{draft}: n=66; \condition{revision}: n=61).  Finally, all participants provided informed consent at the start of the study.

\subsection{Data Collection}
\label{sec:datacollection}

In addition to collecting survey data and the final essay written, we collected: submitted outline, submitted draft, text snapshots at 30-second intervals, keystroke logs, time spent per stage, AI support usage metrics, and AI API requests and responses.




\subsection{Presurvey Measures}

Participants (\(N=253\)) completed a presurvey covering demographics and prior AI use. 
The sample’s mean age was \(38\) years (\(SD=10\)); \(99\%\) identified as native English speakers (250/253). 
Prior AI-for-writing frequency was: Almost every day \(22\%\) (55/253), about once a week \(39\%\) (99/253), about once a month \(10\%\) (26/253), less than once a month \(15\%\) (39/253), and never \(13\%\) (34/253). 
Groups were comparable at baseline: no reliable differences by age (Kruskal–Wallis, \(p=.654\)), native-English status (Pearson’s \(\chi^2\), \(p=.263\)), or writing confidence (Kruskal–Wallis, \(p=.350\)); prior AI-use frequency varied modestly across arms (\(\chi^2\), \(p=.026\)).

We administered four Likert-type psychometric scales (Need for Cognition, Technology Acceptance, AI Writing Efficacy, Writing Self-Efficacy) and averaged item scores per scale. 
Condition-level means (reporting Cronbach’s \(\alpha\) from the overall sample) were similar, with no significant differences between conditions for any of the psychometric scales. Means, standard deviations, and \(\alpha\) are reported in \autoref{sec:surveyqs}. We used these measures to check baseline balance and to describe the sample. To keep our analysis interpretable and centered on stage-of-assistance effects, we do not include these in our analysis.


\section{Planned Analysis}

\subsection{Primary Outcomes}

\paragraph{Psychological ownership:}
Our first outcome is participants' \emph{sense of ownership over the essay}, measured post-task with a single 7-point Likert item (\(1=\) strongly disagree, \(7=\) strongly agree): \emph{``I feel this piece of writing is truly mine.''} We refer to this item as \textit{own\_primary} (higher values indicate greater ownership). For convergent validity we also collected companion items tapping related facets—e.g., \emph{``I feel this writing reflects my own voice and ideas,''} \emph{``I feel I had complete control over the writing process,''} \emph{``I feel that I actively chose all the arguments in this essay,''} and \emph{``I put a lot of effort into writing this essay.''} Analyses for RQ1 use \textit{own\_primary} as the primary outcome; patterns are substantively unchanged when using the companion ownership/agency items.


\paragraph{What participants credit to AI:}
To understand \emph{where} assistance shows up in their work, we asked participants via a post-study survey to split credit between themselves and the AI along two dimensions: (1) idea generation and (2) wording/surface text. These judgments let us separate perceived help with what to say from help with how it is said.
Participants responded on two 0–100 sliders framed as ``What share of the \emph{ideas} in your essay came from \emph{you}'' and ``What share of the \emph{text/wording} came from \emph{you}?'' We reverse-code these to express \% \emph{attributed to AI}, yielding two primary variables that were used to answer RQ2: \textit{ideas\_AI} and \textit{text\_AI}. For analysis we treat them as proportions in \([0,1]\) (with tiny numerical guardrails to avoid exact 0/1) but report results as percentages for readability. We analyze both the components to characterize how each stage of assistance changes perceived contribution.


\paragraph{Essay quality:}
We score writing quality on a 0–9 scale using three rubric dimensions adapted from the IELTS Writing Band Descriptors \cite{IELTSBand}: (i) \emph{Quality of argument}, (ii) \emph{Coherence and cohesion}, and (iii) \emph{Lexical resource and grammar}. An expert English language teacher hand-graded a random 20\% subset of essays ($n = 50$) with this rubric. 

We then developed an automated grading pipeline using OpenAI's \texttt{o4-mini}, a reasoning-oriented language model run as a \emph{separate model instance} from the \texttt{GPT-4o} model that provided AI writing support during the study. The grading prompt instructed the model to act as an expert IELTS examiner, included the full IELTS band descriptors, and supplied 30 human-graded essays as few-shot calibration examples (see \autoref{app:essay_grading} for the IELTS Band Descriptors, provided to the human and LLM grader, and the full LLM prompt).

To evaluate this method, we computed Cohen's $\kappa$ on a held-out set of 20~human-graded essays using a tolerance-based approach: for a given tolerance $\pm t$, a human--LLM score pair was counted as agreement if $|\text{human} - \text{LLM}| \leq t$, and $\kappa$ was then calculated from the observed agreement rate against chance.
This achieved near perfect  agreement  (Cohen’s kappa $\kappa = 0.85$).
Based on this validation, we graded all essays using our method. Our composite quality outcome variable (\textit{quality$\_$overall}) is the mean of the three rubric subscores (range 0–9; higher is better).  Our analysis for RQ3 models this composite outcome while controlling for final word count and total writing time. 

\subsection{Predictors and Covariates Used in Models}

\paragraph{Primary predictor (stage of AI writing assistance):}
All models include the experimental factor \textbf{condition} with four levels:
\condition{none}, \condition{plan}, \condition{draft}, and \condition{revision}.
This factor captures the writing stage at which assistance was offered and is the main variable of interest across RQ1--RQ3.

\paragraph{Task controls:}
To absorb prompt-specific difficulty and style differences, we include a fixed effect for the assigned \textbf{prompt variant} (six levels; e.g., a1--b3) in every model. This keeps comparisons focused on stage-of-assistance rather than writing topic.

\paragraph{Effort covariates:}
Because essay quality can improve with more time and words regardless of assistance, the RQ3 models additionally control for two effort proxies: \textbf{final word count} and \textbf{total time-on-task} in minutes. We intentionally \emph{do not} include these covariates in the ownership or attribution analyses (RQ1/RQ2), as time and length are plausibly downstream of the experimental manipulation.



\subsection{Usage Metrics}

\paragraph{Writing and editing activity:}
We calculate a variety of metrics to measure writing and editing activity.
These include word count of final essay, total keystrokes, time on task, and paste and backspace events.
To characterize how much participants revised their draft, we report a 3-gram and 5-gram word overlap metric for \textit{outline $\to$ draft} and \textit{draft $\to$ final} that indicates how much text is retained between stages. These quantities are reported to illustrate editing dynamics, not used as model covariates.

\paragraph{AI-tool engagement traces:}
For planning support, we report whether or not participants ever requested AI ideas, how many AI ideas were expanded into full outlines, how many AI suggestions (idea or full outline) were pasted into the textbox, how many characters from AI suggestions were pasted into the textbox, and the percent of final characters in the submitted outline that came from AI suggestions.\footnote{Due to logging errors, the exact number of requested ideas is not reported; instead we report only whether a participant ever requested any AI ideas.}
For revision support, we report number of suggestions shown, number applied, acceptance rate, number of characters in final essay that came from revision suggestions, percent of final essay characters that came from revision suggestions, and we also report whether participants ran any revision tool at least once (binary), operationalized as showing at least one suggestion.
These telemetry variables help explain \emph{how} assistance was used within each stage.

\subsection{Model Details}
We fit simple, stage-indexed models for each research question, adding fixed effects for the essay prompt to absorb essay topic differences. Where appropriate, we report condition means (marginalized over prompts) and a small set of prespecified contrasts that capture the comparisons of interest across all outcomes: 
\begin{itemize}
    \item \textbf{C1: No AI vs. AI average},
    \item \textbf{C2: AI Draft vs. the other AI stages},
    \item \textbf{C3: AI Plan vs. AI Revision}.\footnote{We report statistical contrasts (rather than ``eyeballing'' mean differences) because they test the precise patterns we are interested in, pool information across cells, and quantify uncertainty while accounting for prompt effects and unequal cell sizes. Simple visual comparisons can mislead when group sizes or prompts differ.}
\end{itemize}

We focus on these three theory-driven contrasts rather than exhaustively testing all pairwise differences because they map directly onto our research aims, keep interpretation crisp, and avoid unnecessary multiple testing. 
In what follows, $i$ indexes participants, $c_i \in \{\textsc{no-ai}, \textsc{ai-plan}, \textsc{ai-draft}, \textsc{ai-revision}\}$ indicates the assigned condition, and $p_i$ indexes the prompt variant. We use \textsc{no-ai} and a reference prompt $p_{\text{ref}}$ (prompt a1) as baselines for estimation.

\vspace{4pt}
\paragraph{\bf RQ1: Ownership.}
Self-reported primary ownership is treated as approximately continuous on a 1--7 scale. We estimate a linear model of participants’ ownership ratings with stage condition indicators and prompt fixed effects, using robust standard errors.

For reporting, we convert the fitted model into \emph{prompt-adjusted} condition means by averaging predictions equally across prompt variants. We then evaluate three \emph{pre-specified} contrasts on these prompt-adjusted means: \textbf{C1} tests whether introducing \emph{any} AI support shifts ownership relative to \textsc{no-ai} (comparing \textsc{no-ai} to the average of the three AI conditions); \textbf{C2} tests whether \textsc{ai-draft} differs from the average of the other two interactive AI stages; and \textbf{C3} directly compares \textsc{ai-plan} to \textsc{ai-revision}. 

\paragraph{\bf RQ2: Attribution to AI (ideas vs.\ text).}
For each attribution dimension (\textit{ideas\_AI} and \textit{text\_AI}), we treat participants’ slider responses as proportions in $[0,1]$ and fit a generalized linear model appropriate for proportional data with a logit link, allowing extra variability in these percentage judgments. The model includes the same stage indicators and prompt fixed effects as in RQ1, and inference uses robust standard errors.

While reporting results, we back-transform fitted values to the response scale and present prompt-adjusted marginal means for each stage (i.e., expected percent attributed to AI after averaging equally over prompts), with 95\% CIs. We then apply the same three planned contrasts (C1--C3) separately to \emph{ideas} and to \emph{text}, expressing differences in \emph{percentage points} for interpretability.

\paragraph{\bf RQ3: Essay Quality}
We model essay quality as a continuous outcome (0--9), using the same stage factor and prompt fixed effects as in RQ1, plus two covariates capturing task effort: final word count and total time on task. This lets us compare stages after accounting for differences in how much participants wrote and how long they worked. Inference uses heteroskedasticity-robust standard errors.

Our results present \emph{adjusted} (marginal) means by stage---predicted quality averaging equally over prompts and holding the numeric covariates at their sample means---and evaluate the same three pre-specified contrasts (C1--C3) directly on the 0--9 quality scale.

\paragraph{Estimation and reporting:}
All models include prompt fixed-effects to absorb essay-topic differences. We report two-sided 95\% confidence intervals and $p$-values for the small set of prespecified contrasts, which are reused consistently across RQ1--RQ3. Analyses were conducted in \textsf{R} using \textit{sandwich} and \textit{emmeans}. Full model equations, exact contrast weights, and additional estimation details are provided in Appendix~\ref{app:model-details}.

\subsection{Qualitative Analysis} 

We perform a qualitative analysis to add to our understanding of our quantitative results. In particular, we conduct a thematic analysis \cite{braun2012ThematicAnalysis} over the free response questions from the survey, focusing on how the responses may help us understand our three research questions. We had three optional free response questions. The average response length per participant (that is, summing responses to all three questions) was 296 characters, with 87\% of participants answering at least one question, indicating good engagement with the questions. To perform the analysis, two authors read all answers to the free response questions independently, and each created an initial set of themes. Then the two authors came together to discuss the themes and create a single codebook that contained theme names, short descriptors, and relevant quotes. The codebook can be found in \autoref{app:thematic_analysis} and was used to understand and interpret the quantitative results.


\section{Results}

\begin{table}
\caption{Editing telemetry by condition; we report mean (SD). Word count reports words in final essay. Other metrics report amount across all stages (planning, drafting, and revision). Backspace \% is backspaces divided by all key events.}
\label{tab:telemetry_by_condition}
\centering
\small
\begin{tabular}[t]{lcccc}
\toprule
Metric & \textsc{no-ai} & \textsc{ai-plan} & \textsc{ai-draft} & \textsc{ai-revision}\\
\midrule
Word count & 344 (96) & 341 (106) & 342 (45) & 387 (94)\\
Keystrokes & 3281 (1269) & 3284 (1119) & 1144 (600) & 3049 (1229)\\
Time (mins)& 25.1 (5.8) & 26.6 (6.5) & 13.7 (5.1) &26.0 (6.1) \\
Backspace \% & 12.4 (14.0) & 12.0 (11.8) & 12.8 (6.6) & 12.2 (11.8)\\
Paste events & 0.20 (0.48) & 1.73 (7.77) & 0.32 (1.04) & 0.70 (1.49)\\
\bottomrule
\end{tabular}
\end{table}

\begin{table}
  \centering
  \caption{K-gram source–retention by stage. Values are mean\,[SD] \% after requiring both source and target texts to have $\geq 50$ characters. 5-gram indicates near-verbatim carryover; 3-gram captures phrase-level reuse. 
  }
  \label{tab:retention_panels}
\begin{tabular}{lcccc}
  \toprule
   & \textsc{no-ai} & \textsc{ai-plan} & \textsc{ai-draft} & \textsc{ai-revision} \\
  \midrule
  \small{\textbf{Outline  }}\\
  \small{\textbf{$\to$ Draft}} \\
  5-gram & 47.6 [18.2] & 45.7 [21.3] & 48.2 [8.5]  & 50.8 [24.8] \\
  3-gram & 72.4 [11.0] & 70.9 [12.8] & 77.0 [6.0]  & 73.8 [14.8] \\
  \midrule
  \small{\textbf{Draft }}\\
  \small{\textbf{$\to$ Final}} \\
  5-gram   & 93.4 [9.5]  & 95.3 [5.9]  & 88.4 [16.3] & 93.4 [7.6]  \\
  3-gram   & 96.2 [6.0]  & 97.3 [3.8]  & 93.4 [8.9]  & 96.7 [4.4]  \\
  \bottomrule
\end{tabular}
\end{table}

We first describe the usage metrics, which give context to how participants completed the writing task and, when available, used the AI support. We note any relevant differences between the conditions. We then analyze each research question in turn (focused, respectively, on ownership, idea and text attribution, and essay quality), reporting first the quantitative results and then the qualitative results which help us understand the quantitative findings.

\subsection{Descriptive Usage Metrics}

\begin{figure*}
    \centering
    \begin{subfigure}[c]{0.35\textwidth}
        \centering
        \includegraphics[width=\linewidth]{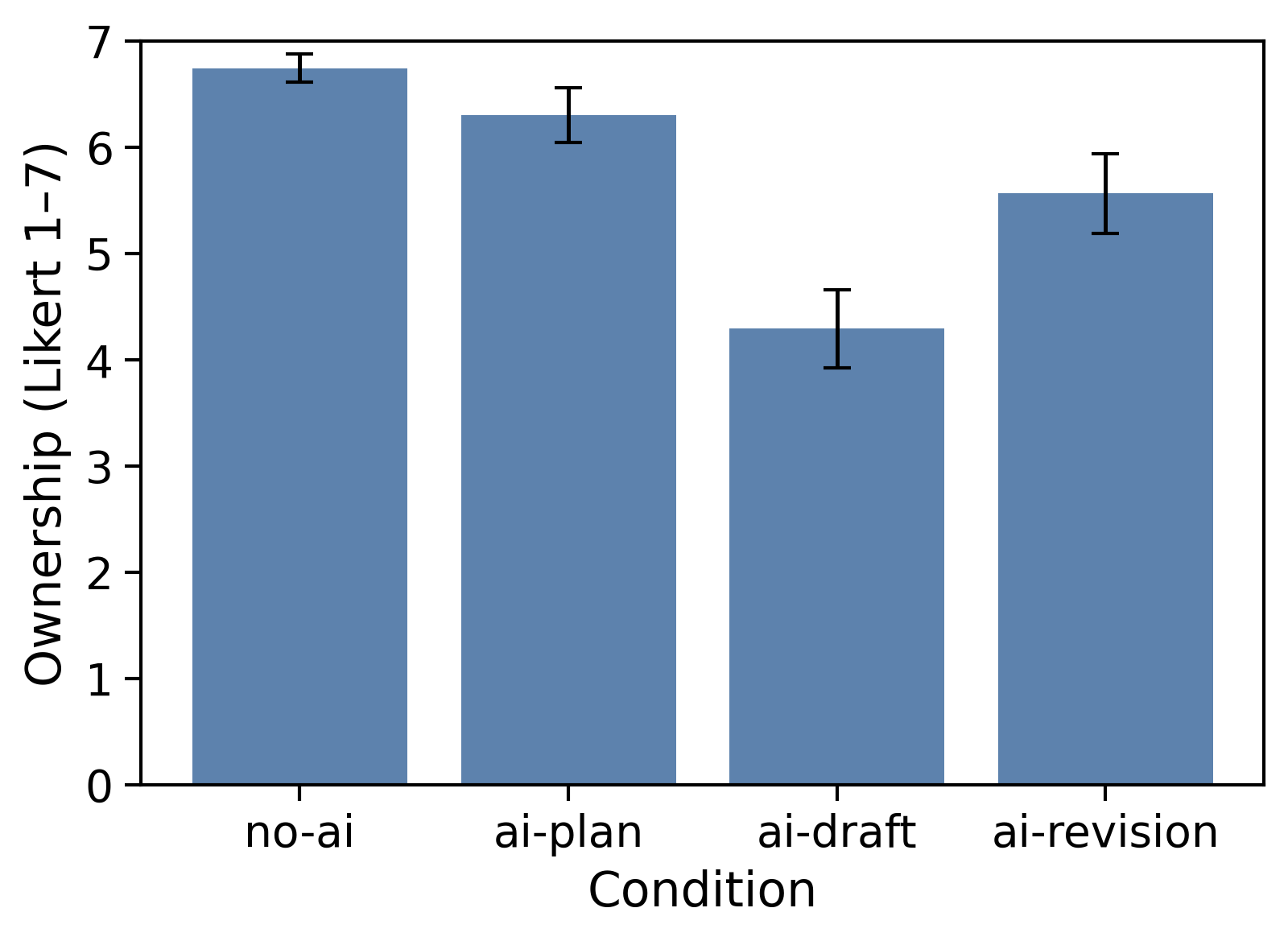}
        \caption{Self-reported ownership by conditions; we show adjusted means and 95\% confidence intervals.}
        \label{fig:rq1results}
        \Description{Bar plot of ownership (likert scale 1-7) for each condition (no ai, ai plan, ai draft, ai revision). Means are 6.74, 6.30, 4.28, 5.57. These results with confidence intervals are reported in \autoref{tab:consolidated_means_by_rq}.}
    \end{subfigure}
    \hfill
    \begin{subfigure}[c]{0.6\textwidth}
        \centering
        \includegraphics[width=\linewidth]{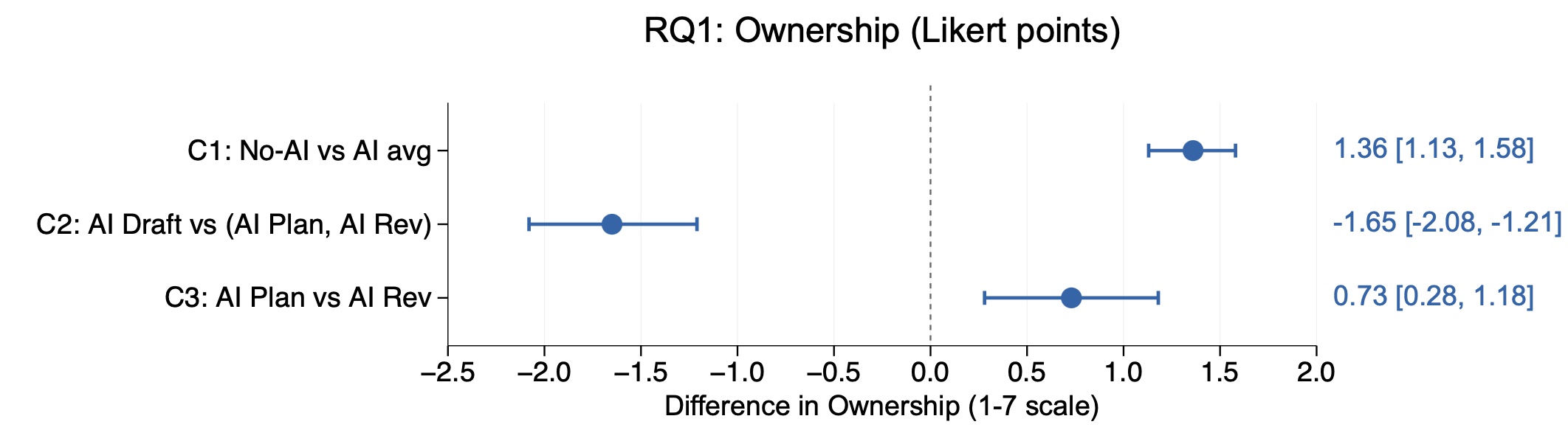}
        \caption{Contrast results: points represent estimated differences with 95\% confidence intervals. All three comparisons are statistically significant ($p<0.05$). The vertical dashed line at zero represents no difference between conditions. Negative values indicate reduced ownership.}
        \label{fig:rq1contrasts}
        \Description{Forest plot of contrast results. C1: +1.36 (CI 1.13, 1.58). C2: -1.65 (CI -2.08, -1.21). C3: +0.73 (0.28, 1.18).}
    \end{subfigure}
    \caption{Results for RQ1. Any AI support decreases ownership compared to a No-AI baseline, with the AI-generated draft decreasing ownership the most, and AI support for revision decreasing ownership more than AI support for planning.}
    \label{fig:rq1all}
\end{figure*}

Our usage metrics indicate that participants largely behaved as expected across the conditions. In \condition{plan}, participants had a similar word count, number of keystrokes, and time on task to the baseline, with a higher number of paste events (pasting AI ideas into their outline). Since word count and keystrokes are dominated by the drafting stage, this indicates that AI support for planning did not drastically change participants' drafting behavior. In \condition{draft}, participants had a similar word count but substantially fewer keystrokes and lower time on task than the baseline, which makes sense given that they did not have to type out a complete draft and skipped the 15 minutes allotted for drafting. In \condition{revision}, participants had a similar time on task but higher word count and fewer keystrokes compared to the baseline, as they pasted in AI revision suggestions which allowed them to add more words to their essay with fewer keystrokes.
All counts can be found in \autoref{tab:telemetry_by_condition}.

Focusing on how the text evolved throughout the task, \autoref{tab:retention_panels} shows how much text was retained between writing stages, tracking changes at the level of 5- and 3-word strings (5-grams and 3-grams). Changes are largely similar across all conditions, with the exception of \condition{draft}: for \textit{Outline $\to$  Draft}, there is more phrase-level reuse (higher 3-gram retention) from the outline in the draft, and for \textit{Draft $\to$ Final}, where there is less verbatim (5-gram) and phrase-level (3-gram) reuse, indicating slightly more change to the text during revision. (We report retention; lower numbers mean more change.) Matching our expectation, participants revised the AI draft more than when they wrote their own draft. 

We also checked how people engaged with the AI support. In \condition{plan}, 65\% of participants used the AI support, with an average of 101 characters inserted into the outline from the AI suggestions. In \condition{revision}, 84\% of participants used the AI support, with an average of 658 characters inserted into the final essay from the AI suggestions. Additional details can be found in \autoref{app:results}, \autoref{tab:engagement_by_stage}. Since not all participants engaged with the AI support, we ran our statistical analysis (reported below) for all participants per condition and then again removing participants who did not use the AI support provided.

\subsection{RQ1: Impact of Stage-Specific AI Support on Psychological Ownership and Agency}
\label{sec:results-ownership}

\paragraph{Quantitative results.}
Psychological ownership was highest with no AI support and declined when AI was introduced, with the size of the decline depending strongly on \emph{where} assistance appeared in the writing process; \autoref{fig:rq1results} shows these results (including confidence intervals). 
On the 1–7 Likert scale, \condition{none} was highest ($\hat\mu=6.74$).
Ownership decreased with AI assistance and followed a clear gradient by stage: \condition{plan} remained relatively high (6.30),
\condition{revision} was lower (5.57),
and \condition{draft} was lowest (4.29).
Adjusted means can be read as the expected average ownership for a stage after accounting for which essay prompt a participant wrote about. 


We then applied the three pre-planned stage contrasts to these prompt-adjusted means to test for significance; see Figure~\ref{fig:rq1contrasts}. The contrast \textbf{C1} (\textsc{no-ai} vs. the average of all AI stages) was $+1.36$ Likert points ($p<.001$), indicating no AI support had significantly higher ownership than when any AI assistance is introduced. \textbf{C2} (\textsc{ai–draft} vs. the average of \textsc{ai–plan} and \textsc{ai–revision}) was $-1.65$ points ($p<.001$), showing that assistance during drafting lowers ownership substantially more than assistance during planning or revision. 
\textbf{C3} (\textsc{ai–plan} vs. \textsc{ai–revision}) was $+0.73$ points ($p=.0016$), indicating planning preserves more ownership than revision. Taken together, the contrast estimates reinforce that \emph{where} AI is placed matters: drafting produces the largest displacement in ownership, while planning softens that trade-off. Restricting to participants who actively engaged with the tools yields the same pattern, with the \condition{plan} versus \condition{revision} contrast becoming slightly more pronounced (Appendix~\ref{app:robustness-engagement}).

We also examined participants' sense of control (``\textit{I had complete control over the writing process}'') as a secondary outcome. The directional pattern mirrored ownership—highest in \condition{none}, with \condition{plan} showing the smallest reduction among AI stages, \condition{revision} intermediate, and \condition{draft} the largest decrease, thereby reinforcing the robustness of our results.

Our survey collected two agency items (``\textit{I feel I had complete control over the writing process}'' and ``\textit{I feel I actively chose all the arguments in this essay}''). These two measures were moderately correlated ($\rho\!\approx\!.67$; $\alpha\!\approx\!.68$). Also, agency correlated with ownership at a moderate level ($\rho\!\approx\!.47$–$.70$), indicating related but non-redundant constructs. Re-estimating the RQ1 model with a two-item agency composite produced the same stage ordering as ownership (\condition{plan} highest among AI stages, \condition{draft} lowest, \condition{revision} in between) with comparable effect sizes. 

Looking at the editing metrics in \autoref{tab:telemetry_by_condition} may explain some but not all of these results. Participants in the \condition{draft} condition spent less time on task and typed fewer keystrokes, which could explain at least part of the drop in ownership. This aligns with the labor or effort "route" to ownership. On the other hand, the \condition{plan} and \condition{revision} conditions seem quite similar to the \condition{none} condition in these editing metrics yet have different ownership levels, confirming that labor or effort is not the only path that impacts ownership feelings.

Self-reported effort (``\textit{I put a lot of effort into writing this essay}'') correlated positively with ownership ($\rho\!\approx\!.57$). 
Self-reported effort was high in \condition{none} $(6.53 \pm 0.79)$, \condition{plan} $(6.35 \pm 0.89)$, and \condition{revision} $(6.30 \pm 0.84)$, but lower in \condition{draft} $(4.58 \pm 1.69)$, mirroring editing activity.
Across the full sample, effort correlated positively with keystrokes $(\rho = .25)$, 
consistent with keystrokes indexing motor activity while much of effort is cognitive.

\begin{figure*}
    \centering
    \begin{subfigure}[c]{0.35\textwidth}
        \centering
        \includegraphics[width=\linewidth]{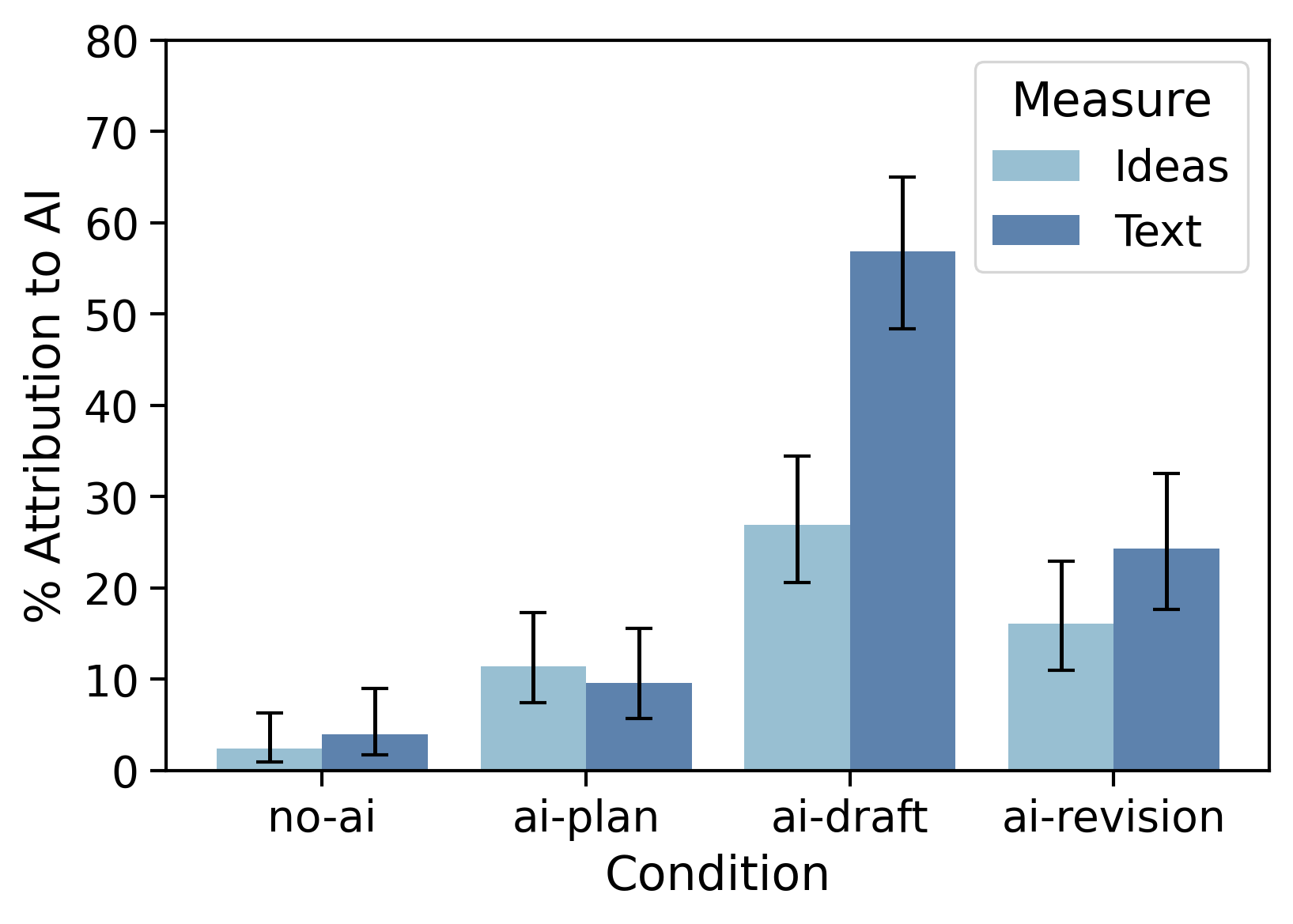}
        \caption{\% attribution of text and ideas to the AI; we show adjusted means and 95\% confidence intervals.}
        \label{fig:rq2results}
        \Description{Bar plot of text and idea attribution to AI for each condition (no ai, ai plan, ai draft, ai revision). Means are (2.4, 4.0), (11.4, 9.6), (26.9, 56.9), (16.1, 24.3). These results with confidence intervals are reported in \autoref{tab:consolidated_means_by_rq}.}
    \end{subfigure}
    \hfill
    \begin{subfigure}[c]{0.6\textwidth}
        \centering
        \includegraphics[width=\linewidth]{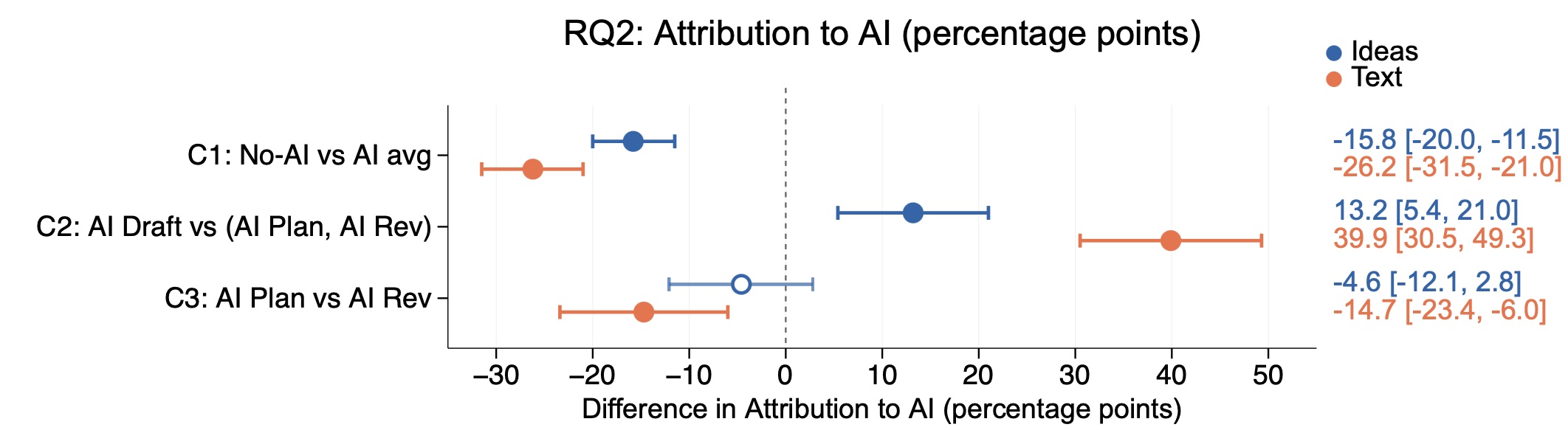}
        \caption{Contrast results: points represent estimated differences with 95\% confidence intervals. Filled circles indicate statistically significant effects ($p<0.05$); hollow circles indicate non-significant effects. The vertical dashed line at zero represents no difference between conditions. Positive values indicate greater attribution to AI.}
        \Description{Forest plot of contrast results for ideas and text attribution to AI. C1: (-15.8, -26.2). C2: (+13.2, +39.9). C3: (-4.6, -14.7).}
        \label{fig:rq2contrast}
    \end{subfigure}
    \caption{Results for RQ2. Attribution of text and ideas to AI tracks with our ownership results, with highest ownership aligning with lowest attribution to AI. Attribution of ideas and text mostly track across conditions, with the biggest discrepancy in the AI-draft condition.}
    \label{fig:rq2all}
\end{figure*}

\paragraph{Qualitative results.}

The thematic analysis of the free response questions can help interpret the quantitative results. Participants talked about effort and investment, consistent with our quantitative results. As one participant wrote, "\textit{I wrote all of it by myself, and am proud of the work I did}" (\condition{none}), while another wrote, "\textit{I put effort into writing my personal views about genetic editing}" (\condition{plan}). 

However, participants also talked about emotional resonance, self-efficacy, and their sense of voice or expression. This came up a lot in the \condition{none} condition, where participants talked about writing "\textit{from the heart}", how the topic is one they "\textit{hold dear}" or one they are "\textit{extremely passionate about}". Participants across all conditions talked about whether or not the essay was in their "\textit{voice}" as a factor in ownership. As one participant in the \condition{draft} condition wrote, "\textit{The outline was mine, but the voice, the writing style, was not}". 
These reflect the affective side of ownership: how ownership relates to feelings of identity and investment, rather than effort or control. Participants in all three AI support conditions wrote about how the AI support did or did not change their ideas; we investigate this in the next section.

\subsection{RQ2: How Stage-Specific AI Support Reallocates Provenance: Ideas vs.\ Text}

\subsubsection{Quantitative results.}

Across stages, AI support shifted perceived provenance toward the AI, with the strongest shift when assistance was provided during drafting; see \autoref{fig:rq2all}, which includes confidence intervals.
On \emph{ideas}, participants in \condition{none} attributed on average $2.4\%$ of ideas to AI.\footnote{This small but nonzero baseline is plausible for three reasons: first, some participants appear to interpret ``AI influence'' broadly (e.g., the study’s title and interface cues may have primed AI as a salient frame, and some respondents may include generic algorithmic tools or ambient AI-shaped discourse in their judgments), second, at least two participants seem to have misinterpreted the slider and entered 0 instead of 100, and third, percentage judgments invite small nonzero responses even when influence is minimal. All detected uses of external AI or large copy–paste events were excluded from the analysis, so the baseline does not reflect active AI use in the task.} 
This rose to $11.4\%$ for \condition{plan}, 
$26.9\%$ for \condition{draft}, 
and $16.1\%$ for \condition{revision}.
On \emph{text}, we found the same pattern but amplified: $4.0\%$ for \condition{none}, 
$9.6\%$ for \condition{plan}, 
$56.9\%$ for \condition{draft}, 
and $24.3\%$ for \condition{revision}. 
In other words, with drafting assistance, a majority of the \emph{final text} was perceived as coming from AI, and participants also credited AI with a substantial share of ideas; planning produced modest AI provenance on both dimensions; revision sat in between. 

To quantify these patterns, we applied the three planned contrasts from our model on the response scale (percentage points). As can be seen from Figure~\ref{fig:rq2contrast}, relative to \condition{none}, the average of the three AI stages increased the share attributed to AI by $+15.8$\% for ideas ($p<.001$) and by $+26.2$\% for text ($p<.001$). Drafting differed sharply from the other AI stages: \condition{draft} exceeded the mean of \condition{plan} and \condition{revision} by $+13.2$\% on ideas ($p<.001$) and by $+39.9$\% on text ($p<.001$). Finally, planning and revision were similar on ideas but diverged on text (planning $-$ revision: $-14.7$\%; $p=.001$). 

While this pattern is consistent with the AI-generated draft producing the largest amount of text, several results diverge from our expectations. It is notable that only a small percentage of ideas are attributed to AI in the \condition{plan} condition (11.4\%), even though 65\% of participants requested AI ideas. Separating users from non-users clarifies this: among those who requested ideas, attribution was 17.8\%, while non-users stayed near baseline (2.9\%). The stage-level mean is modest because it mixes these groups. Ownership, however, did not differ between users and non-users within \condition{plan}---shifts in attribution don't straightforwardly translate to ownership. (In \condition{revision}, 84\% of participants used the tool, leaving only 10 non-users; splits are directionally consistent but underpowered.) Additionally, an AI-generated draft based on participants' own outline produced \textit{more} ideas than expected (27\%); one might imagine that an outline could contain all the ideas necessary to write an essay but we found this was not necessarily the case. Taken together, these results indicate that the drafting stage is not necessarily one of simply ``filling in'' an outline, but a stage in which ideas are developed. We investigate this further with the qualitative analysis, which gives us insight into how our participants understood this outline-to-draft process.


We can compare these results to our details about n-gram retention (\autoref{tab:retention_panels}). 
The \textbf{draft} to \textbf{final} step shows very high stability: 3-gram retention is $\geq 93\%$ in every condition, and 5-gram retention is $\sim$93–95\% in \condition{none}, \condition{plan}, and \condition{revision}, dipping modestly to $\sim$88\% in \condition{draft}. In other words, people did not revise too much in any condition; when AI provides the draft, people do revise a bit more (lower 5-gram), but the underlying phrases remain largely intact (still high 3-gram). These results imply that, in our experiment, drafts (human written or AI-generated) anchored the final product, and AI support mainly shifted \emph{how verbatim} the carryover was (exact strings vs. rephrased), rather than whether content "flowed" through. 
\begin{figure*}
    \centering
    \begin{subfigure}[c]{0.35\textwidth}
        \centering
        \includegraphics[width=\linewidth]{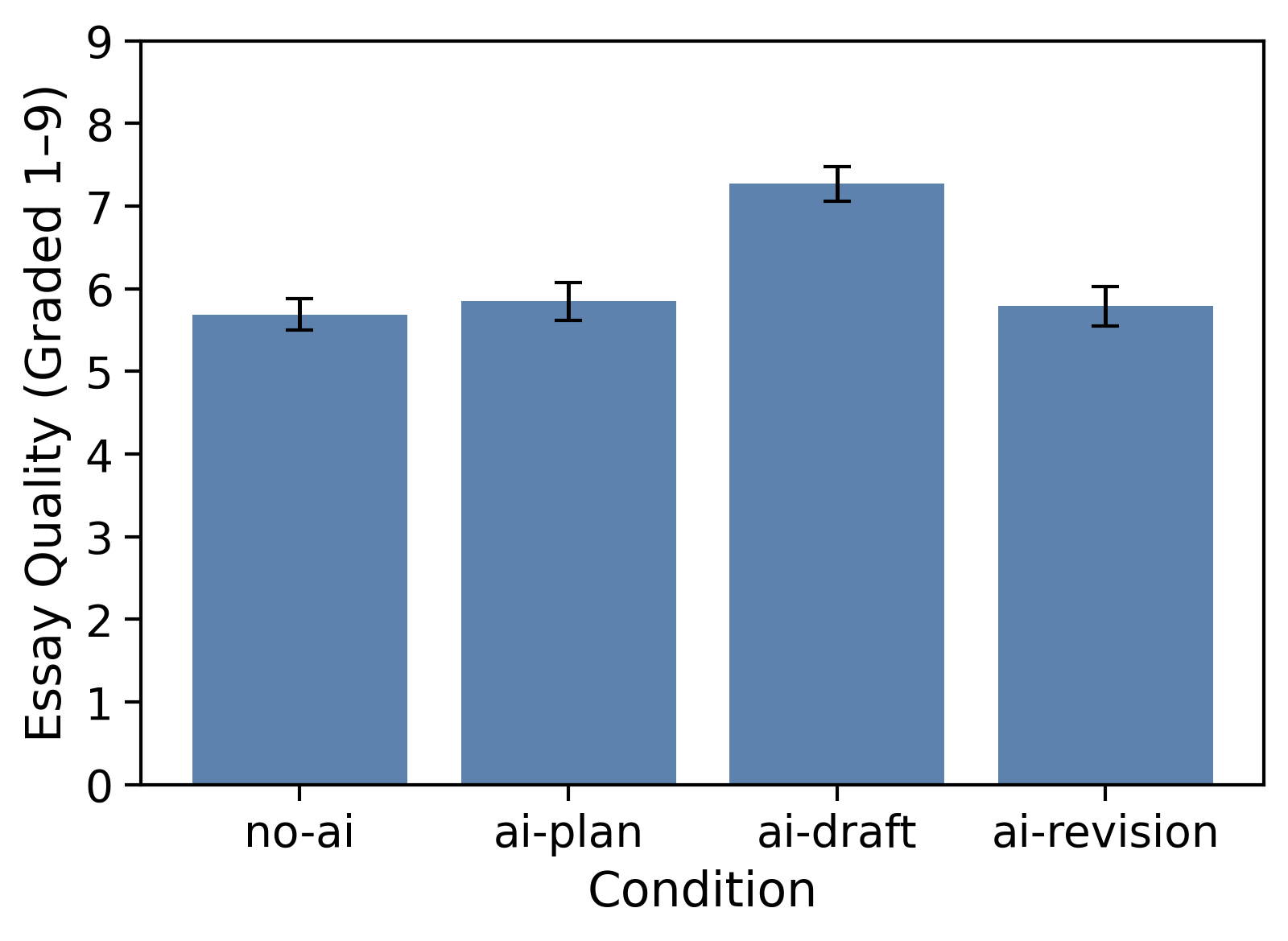}
        \caption{Essay quality by condition; we show adjusted means and 95\% confidence intervals.}
        \label{fig:rq3results}
        \Description{Bar plot of essay quality (graded 1-9) for each condition (no ai, ai plan, ai draft, ai revision). Means are 5.69, 5.85, 7.27, 5.79. These results with confidence intervals are reported in \autoref{tab:consolidated_means_by_rq}.}
    \end{subfigure}
    \hfill
    \begin{subfigure}[c]{0.6\textwidth}
        \centering
        \includegraphics[width=\linewidth]{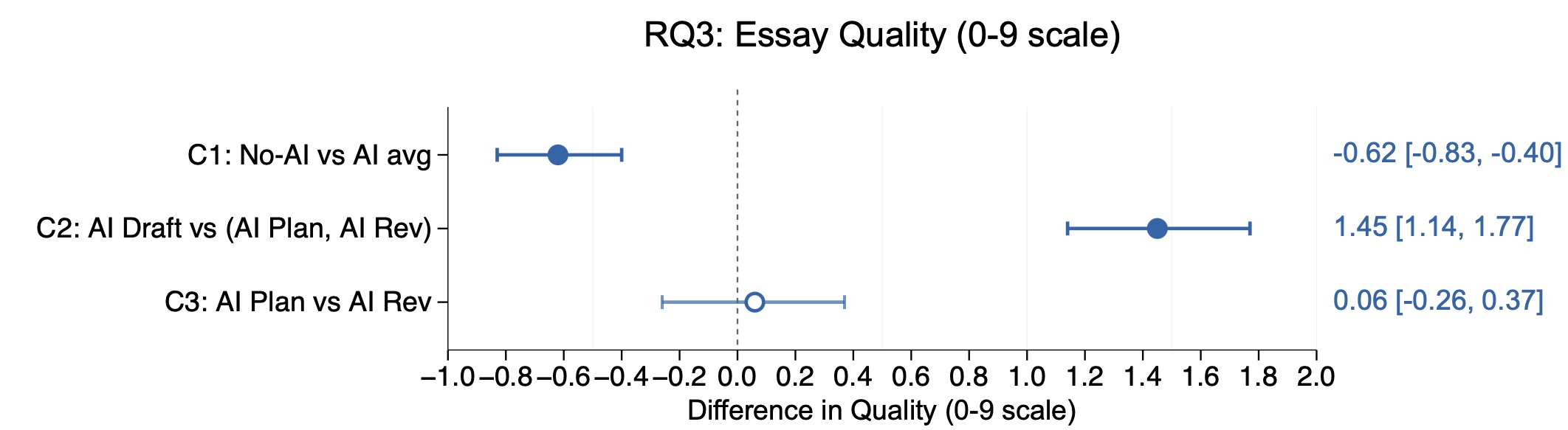}
        \caption{Contrast results: points represent estimated differences with 95\% confidence intervals. Filled circles indicate statistically significant effects ($p<0.05$); hollow circles indicate non-significant effects. The vertical dashed line at zero represents no difference between conditions. Positive values indicate improved quality.}
        \Description{Forest plot of contrasts. C1: -0.62. C2: +1.45. C3: +0.06.}
        \label{fig:rq3contrast}
    \end{subfigure}
    \caption{Results for RQ3; note that our model controls for prompt variant as well as final word count and total minutes on task. AI draft increases essay quality significantly over No AI and other AI support conditions.}
    \label{fig:rq3all}
\end{figure*}

\subsubsection{Qualitative results.}

Participants wrote about the myriad ways in which ideas were accepted, ignored, integrated,  or changed by either themselves or the AI assistance. In the \condition{plan} condition, participants seemed to often request ideas and then not use or incorporate them. In some cases this was because the AI-generated ideas were redundant with participants' own ideas, e.g., "\textit{the AI did come up with outline ideas, but they matched/confirmed what I had already thought}".
In other cases, participants chose to use their own ideas over the AI-generated ideas, e.g., "\textit{the AI gave me a few ideas, but I ended up choosing to stick closely to my own thoughts}".
However, other participants did combine their own ideas with the AI ideas; for instance, one participant wrote, "\textit{About half of the points were mine and half of them were either introduced by the AI}".

In the \condition{draft} condition, participants discussed how the AI-generated draft contributed to their ideas. Participants used phrases like 
"\textit{expanded on}" and  
"\textit{fleshed out}"  
to describe this. Some said that their ideas were accurately reflected in the AI draft, e.g., "\textit{it stuck to my key arguments}".
Others note that the AI draft "\textit{brought up some points I missed}" or "\textit{provided the details such as numbers, statistics, and potential other financial schemes}". Many participants noted that
"\textit{[the essay] doesn't really feel as if it is writing in my own style and voice}". In contrast, one participant wrote that they "\textit{could have written a similar essay}" as an explanation for why they continued to feel ownership over the resulting essay. 
However, the effect on ownership was not always clear, and the line between participant ideas and AI ideas was blurry. As one participant wrote, "\textit{The ideas were my own, but what resulted from those ideas was not}". 

In the \condition{revision} condition, the AI suggestions did seem to introduce new ideas and not just stylistic changes. 
One participant wrote, "\textit{AI did not provide any additional thoughts or arguments, it just provided supplementary evidence to support my comments/thoughts}", suggesting that evidence does not count as additional thoughts. Another wrote, "\textit{Because the ideas are overall still in tact and fully mine... AI just expanded on the direction I was already going}", again suggesting that expanding on ideas does not significantly change them. 
The line between participant and AI ideas was blurry, depending both on the way in which a participant incorporated AI ideas and their perception of how those changes impacted the final essay.

These results point to the complicated nature of writing, which is not merely an act of translating fully formed concepts into text. As one participant in the \condition{draft} condition wrote, 

\begin{quote}
    \textit{I feel as though if I were to write the essay it would take me much longer... When I write essays, the process of writing the essay helps me determine my final decision on the matter, the more I write the more I allow myself to think and reason with the evidence I found.}
\end{quote}


\subsection{RQ3: Effects of Stage-Specific AI Support on Essay Quality}

Across conditions, AI support improved essay quality, with the strongest gains when assistance was provided during \emph{drafting}, see \autoref{fig:rq3all}. (Note that our linear model that controls for prompt variant as well as final word count and total minutes on task, so stage effects are not simply artifacts of writing more or spending more time on the task.) Essays were scored on a 0-9 scale; relative to \condition{none} ($5.69$), quality was similar for \condition{plan} ($5.85$) and \condition{revision} ($5.79$), but markedly higher for \condition{draft} ($7.27$). 

We then applied the three planned stage contrasts to the adjusted means (Figure~\ref{fig:rq3contrast}). First, comparing \textsc{no-ai} to the average of the three AI stages (\textbf{C1}) shows a $+0.62$-point advantage for AI support ($p<.001$). Second, \textbf{C2} isolates drafting from the other assisted stages: \textsc{ai–draft} scored $+1.45$ points higher than the average of \textsc{ai–plan} and \textsc{ai–revision} ($p<.001$). Third, \textbf{C3} finds \textsc{ai–plan} and \textsc{ai–revision} statistically indistinguishable on quality. 
Taken together with the findings in Section \ref{sec:results-ownership}, these results point to a practical trade-off: the stage that most boosts quality (drafting) is also the stage that most depresses psychological ownership, whereas planning preserves ownership with only modest quality change, and revision sits between.

\section{Discussion}

\subsection{Writing as a mode of learning}

Our study intended to bring nuance to the concern that AI writing support decreases writers’ sense of ownership by isolating support to different writing stages. Our findings extend beyond documenting ownership drops from AI support, and find, in line with a long history of writing studies research, that the very process of writing a draft is key to the development of its writer’s final perspective. While people use LLMs to support their writing process in a variety of ways, and some usage may conceivably increase the effort or time dedicated to a writing task, early research on LLM usage for writing indicated its popularity was due to increases in productivity: getting existing writing tasks done faster and at a higher quality \cite{noy2023ExperimentalEvidenceProductivity}. Writing studies research indicates that writing is temporally distinct from speech because of “the possibility of revision” \cite{emig1977WritingModeLearning, sommers1980RevisionStrategiesStudent}, letting writing be a form of discovery rather than simple “telling.” Doing the same writing task in less time conceivably results in less discovery, or the discovery being partially done by the LLM rather than the human writer, which would lead to decreased sense of ownership over the final result. 

Our results confirm this hypothesis. Even when participants got an AI draft based on their own outline, and then revised this draft more than those who wrote their own draft, their ownership was still significantly lower than not only those without AI support, but those who took outline ideas from an AI or took revision suggestions from an AI. Although we may consider it “obvious” that an AI draft would significantly decrease ownership, it is notable that participants also thought the AI draft introduced more ideas than AI outline contributions or AI revision contributions. In fact, whenever participants took fluent text from the AI, participants not only attributed text but also ideas to the AI, suggesting that decoupling ideas from fluent text is difficult.

In certain contexts, efficiency may be key, especially when writing is not intended to clarify one’s thoughts or perspectives. We also found, in line with other work, quality gains as well \cite{dhillon2024ShapingHumanAICollaboration, noy2023ExperimentalEvidenceProductivity}, although that may be an artifact of the task — participants may have needed more time to achieve a similar level of polish, which can be expedited with AI support, and this kind of short-form argumentative writing task is one that AI support excels at \cite{tang2025AreLLMsReady}. Still, we argue that in some cases an implicit utility of writing is becoming more familiar with one’s own thoughts, which inevitably changes and ideally improves them. We must consider in which contexts it is appropriate to speed up writing, and argue that speeding up writing likely incurs an ownership cost.

\subsection{Temporality in writing assistance}

Our study hypothesized that support at different writing stages results in a different drop in ownership. Although the type of support at different stages looks quite different, in our study these stages were also temporally distinct. Participants who took AI ideas during the outlining stage simply had more time writing on their own afterwards. There may have been a recency bias: AI support during outlining is “forgotten” by participants and therefore has less impact on ownership and attribution of text and ideas. Another interpretation is that participants have more time to work with and develop the ideas on their own; even if they took a full outline from the AI support, the time and effort spent developing it into a draft and through revision results in their investment of effort (material ownership) and investment of ideas (idealist ownership). 

This is in contrast to work on idea generation and AI support. \citet{qin2025TimingMattersHow} find that having LLMs produce ideas \textit{after} the participant has produced ideas increases self-credit, finding that autonomy is the mediator for ownership, and early independent ideation increases autonomy. However, they did not have a condition in which participants had to ideate on their own after co-ideation with the AI support, nor did participants have to then work on their ideas in any way. A more comparable study is \citet{si2025IdeationExecutionGapExecution}, where they compare LLM- to human-generated research ideas, and then hire researchers to put 100 hours of work on each idea. This study finds that while LLM and human ideas rank similarly before execution, they perform worse after execution. However, this paper does not study ownership directly; instead, the researchers who perform execution are allowed to take full ownership of any resulting work. 

Our findings also suggest that the psychological effects of stage-specific support depend on actual engagement, not just availability. In \condition{plan}, 35\% of participants never requested AI ideas, and their ownership and attribution scores were indistinguishable from \condition{none}. By contrast, \condition{revision} had much higher uptake (84\%). Planning support may be easier to ignore---writers who arrive with ideas in mind may see little need for AI suggestions, whereas revision tools offer concrete improvements to any existing draft. Designers of AI writing tools should consider not only when to offer support, but how to encourage meaningful engagement, particularly for upstream stages where the value of assistance may be less immediately apparent.

\section{Design Directions}

Based on findings and discussion, we propose three directions for future work on designing AI writing tools. Our findings point to being careful with fluent prose, where direct use of AI text (as opposed to, e.g., paraphrasing) resulted not just in attributing text to the AI but also attributing ideas. Our design implications all investigate how to reduce reliance on AI prose in the final output while preserving other AI support abilities. While certainly there are times to rely on AI prose, we believe integrating AI support without relying on actual text is underexplored but well supported as a valuable research direction.

Recent work on the design of AI writing interfaces has explored many spatial approaches, such as DirectGPT \cite{masson2024DirectGPTDirectManipulation}, Textoshop \cite{masson2025TextoshopInteractionsInspired}, and Nabokov’s Cards \cite{carrera2025NabokovsCardsAI}. \citet{buschek2024CollageNewWriting}, in his paper on collage as a concept for writing tools, argues that AI has re-introduced the fragmentary nature of the avant-garde literary movement focused on found texts and the collaging of these as an alternate approach to writing. We take on these ideas when thinking about design implications, and consider how to also incorporate time as a design concept. We argue that time is just as important as space when designing interfaces.

In these design ideas, we seek to expand the vision of how writing support tools might work, and how we might preserve “writing as a mode of learning” while also making use of a new and powerful tool.


\paragraph{Fade-out AI interactions:} AI support could be phased out as a draft becomes more and more refined. Initially, writers can receive substantial scaffolding—ideas, outlines, whole paragraphs. As the draft develops, AI contributions could become sparser or more fine-grained, shifting from paragraphs to sentences, and eventually to only flagging potential issues. This mirrors natural writing processes where early drafts are exploratory and later revisions are refinement. By decreasing AI visibility over time, such interfaces could encourage writers to increasingly "take over" their work, supporting the gradual psychological transition from external scaffold to owned artifact.

\paragraph{Provide incomplete text that requires rewriting:} AI tools could provide incomplete rather than fluent text: bullet points, rough phrasings, or unpolished sketches that cannot simply be copied and pasted. (A similar idea was explored in \citet{zhou2023creative}.) Writers would be required to rewrite these into proper prose themselves. This approach could be incorporated into collage-like writing interfaces, but disallows direct use of AI-generated sentences. By requiring writers to produce fluent prose, such tools could maintain quality benefits while preserving the discovery that happens through articulation.

\paragraph{Turn the chatbox into a writing space:} Much AI writing assistance is requested via a chat interface, reinforcing an anthropomorphic metaphor of "asking an assistant" rather than engaging with raw material. Drawing on the collage concept \cite{buschek2024CollageNewWriting} and direct manipulation interfaces \cite{masson2024DirectGPTDirectManipulation}, we propose making AI chat outputs directly editable within the conversation itself. Rather than copying AI text into a document, writers could revise, reorganize, and rewrite AI responses in place before incorporating them. This shifts AI contributions from finished deliverables to malleable drafts, more like word processor affordances that increased line-level editing. Making AI output editable could reduce the psychological distance between AI contributions and the writer's own work, supporting a more iterative, hands-on relationship with AI-generated content.

\section{Limitations and Future Work}

\paragraph{Engagement in writing task.} We ran our study remotely, with online participants recruited via Prolific; we had to reject 18\% of all submissions due to unapproved, external AI usage. 
We incentivized participants via bonuses for high quality writing, and answers to the free response questions indicated at least some participants enjoyed the task and engaged deeply, but controlled studies cannot replicate the kind of authentic engagement found in the wild. Our task was a standard one, but one with no clear stakes; people might be more hesitant to accept AI suggestions if stakes are higher, for instance if they may get in trouble for inaccuracies or are trying to perform at a high level in order to get promoted. Relatedly, although we did look at attribution we did not look at the related construct of responsibility. Future work could consider more analytical, high stakes, or personally motivated writing tasks to understand how that may change AI interactions.

\paragraph{Differences between writing stages and type of support.}
Writing is not a linear process: while people do plan, draft, and revise, they do not always do so in exactly that order, and are likely to go through multiple rounds of planning, drafting and revision \cite{flower1981CognitiveProcessTheory}.
Forcing our participants to outline, draft, and then revise was somewhat artificial and thus so were our AI support designs; e.g., participants were only allowed to generate an AI draft once rather than have a multi-turn conversation.\footnote{We did consider allowing participants to regenerate their draft, which is also a common usage pattern; however, we worried that this may have spilled into revision, making it difficult to compare the stages. To preserve our focus on temporal stages, we wanted to isolate the drafting support.} This could have influenced our findings for the drafting stage; if participants had more options for engaging in this stage they may have felt more ownership over the text and ideas. That said, it remains difficult to decouple drafting from planning and revision, which is why we opted for a linear set of stages, and allowing more options in the AI draft condition may have reduced participants' engagement in revision making our comparisons invalid. Future work could give people access to AI at different temporal points in their writing process (e.g., in the first, middle, or last 10 minutes of a 30 minute writing session) without restricting what kind of writing they can do, and post-hoc analysis could determine the kind of AI support they opted for. This would make for a stage-like measure, where we control \textit{when} in the writing process people use AI but not \textit{what kind} of support they request, which would deepen our understanding of how AI support changes through the temporally-complex writing process.

\paragraph{Qualities of the writing task.} No one writing task will be representative of all writing tasks. While we found writing quality gains with increased AI support, in line with prior work \cite{dhillon2024ShapingHumanAICollaboration}, other work has found that quality gains are uneven and depend on the specific writing task at hand \cite{dellacqua2023NavigatingJaggedTechnologicala}. Additionally, our writing task is one that AI is known to excel at \cite{tang2025AreLLMsReady}, and our grading scheme (although internationally standard) may be biased towards stylistic fluency; our human and LLM graders may be similarly biased. Many writing instructors can critique the task as an inauthentic rhetorical situation performed under artificial time constraints. Our work should be replicated in other writing tasks, genres, and time frames that may be more difficult for AI tools to perform well on, particularly with regard to ownership, personal investment, and connection. Some studies on integrating LLMs into writing processes have had participants write analysis of provided data \cite{umarova2025HowProblematicWriterAI} or personal narratives \cite{schnitzler2026amplifying}, tasks that engage with authentic and concrete rhetorical situations and/or have stronger personal or professional stakes for writers. Finally, professional writers likely respond to AI support very differently from amateur writers, and AI support likely impacts writing quality differently.

\section{Conclusion}

In this paper we investigated how AI support at different writing stages---planning, drafting, and revision---impacts feelings of ownership. We designed AI support tools at each of these stages to mimic typical AI support for writing found in the wild. We ran an online, between-subjects experiment in which participants were asked to write a 200-300 word essay. We found that AI support at the drafting stage resulted in the greatest decrease in ownership, with support at the revision stage having the second greatest decrease and planning support having the smallest decrease (compared to a no-AI baseline). This decrease in ownership mapped neatly onto an \textit{increase} in ideas and text attributed to the AI support, and an increase in essay quality. 

\begin{acks}
Thanks to Ray Zhou for support with the grading rubric and the essay grading.
\end{acks}

\bibliographystyle{ACM-Reference-Format}
\bibliography{citations}


\newpage

\appendix

\clearpage
\onecolumn
\section{Essay Writing Prompts}
\label{app:methods}


\begin{table*}[h]
\begin{center}
\begin{tabular}{ l l p{12cm} }
    \hline
    ID & n (\%) & Essay Prompt \\
    \hline
    a1 & 58 (23\%) & Should public schools ban smartphones during the school day, or permit limited use for learning and emergencies?\\
    a2 & 22 (9\%) &Should we bring back extinct species like woolly mammoths using genetic engineering, or leave extinction as a natural boundary that shouldn't be crossed?\\
    a3 & 48 (19\%) & Should large employers require workers to return to the office several days a week, or allow fully remote schedules by default? \\
    \hline
    b1 & 53 (21\%) & Should parents have the right to genetically edit their unborn children to prevent diseases, or should we ban genetic modifications to preserve natural human diversity?\\
    b2 & 33 (13\%) & Should cities ban gas-powered leaf blowers to reduce noise and pollution, or leave equipment choices to homeowners and landscapers?\\
    b3 & 39 (15\%) & Should the federal government broaden student-loan forgiveness programs, or prioritize other ways of addressing education debt?\\
    \hline
\end{tabular}
\caption{List of essay prompts used in the study. Participant selected one prompt from either set A or set B. We report $n$ -- how many participants selected that prompt.}
\label{tab:essaytopics}
\end{center}
\end{table*}

\clearpage

\section{System Design}
\label{sec:screenshots}



\begin{figure*}[!h]
    \centering
    \includegraphics[width=.8\linewidth]{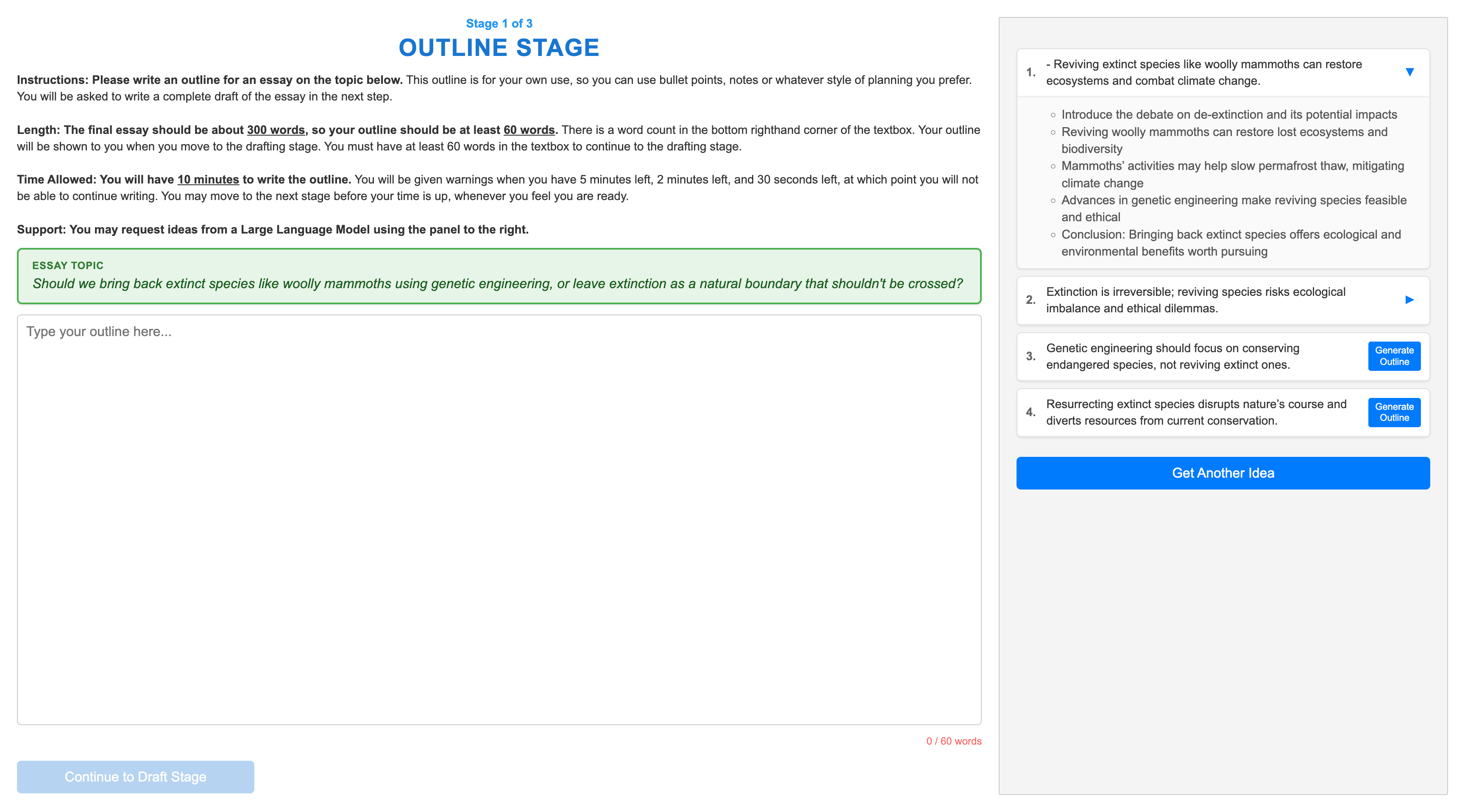}
    \caption{\textbf{Outline Stage:} The AI suggests supporting ideas in bullet points, shown in the right panel. In conditions where AI support is not provided, the right panel is empty.}
\end{figure*}




\begin{figure*}[!h]
    \centering
    \includegraphics[width=.8\linewidth]{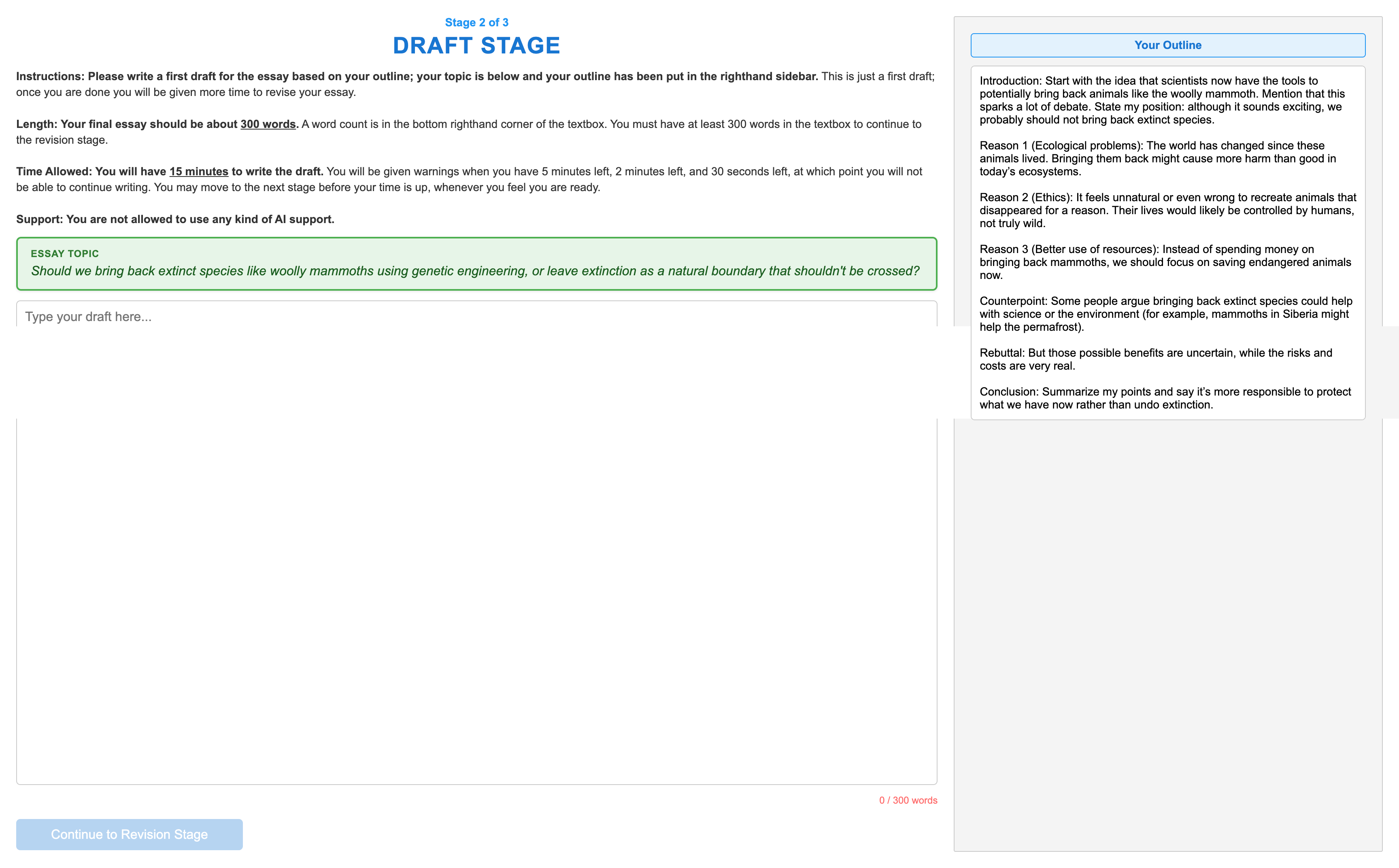}
    \caption{\textbf{Draft Stage:} The user manually drafts their essay based on the outline shown on the right from the previous step. In condition 3 (AI-generated draft) participants skip this stage, and the AI-generated draft automatically populates the textbox in the revision stage.}
\end{figure*}
\clearpage

\begin{figure*}[!h]
    \centering
        \includegraphics[width=.8\linewidth]{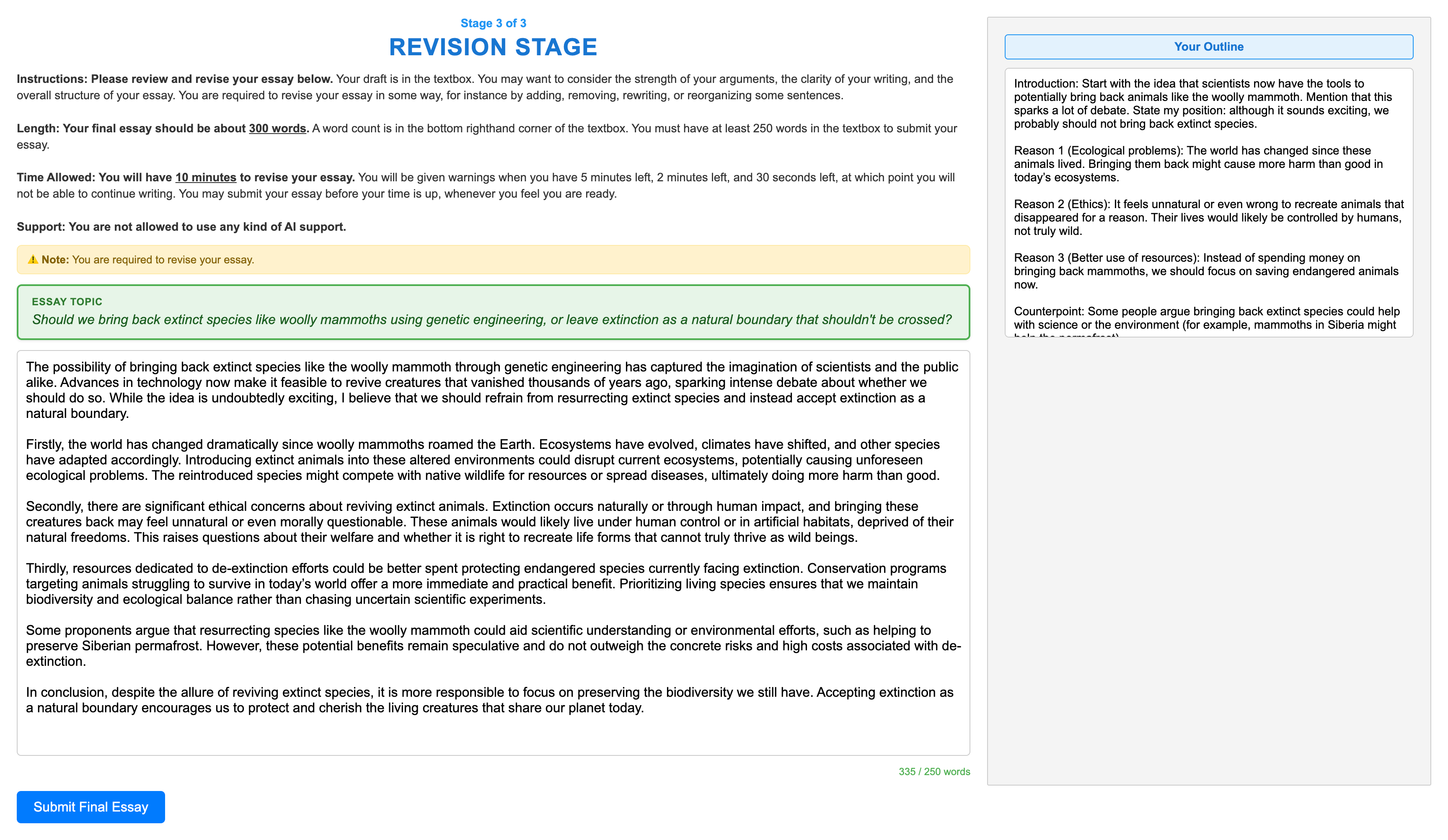}
        \caption{\textbf{Revision Stage:} User revises essay in main textbox. In the no-AI condition, their outline is provided in the right panel, as above. In the AI support condition, the right panel contains the revision tools. These are shown in the next figure.}
\end{figure*}

\begin{figure*}[!h]
    \centering
    \begin{subfigure}{0.32\textwidth}
        \includegraphics[width=\linewidth]{sections/appendix_system_screenshots/revisionAI-1.png}
        \caption{AI Revision Tool: Argument Improver}
    \end{subfigure}
    \hfill
    \begin{subfigure}{0.32\textwidth}
        \includegraphics[width=\linewidth]{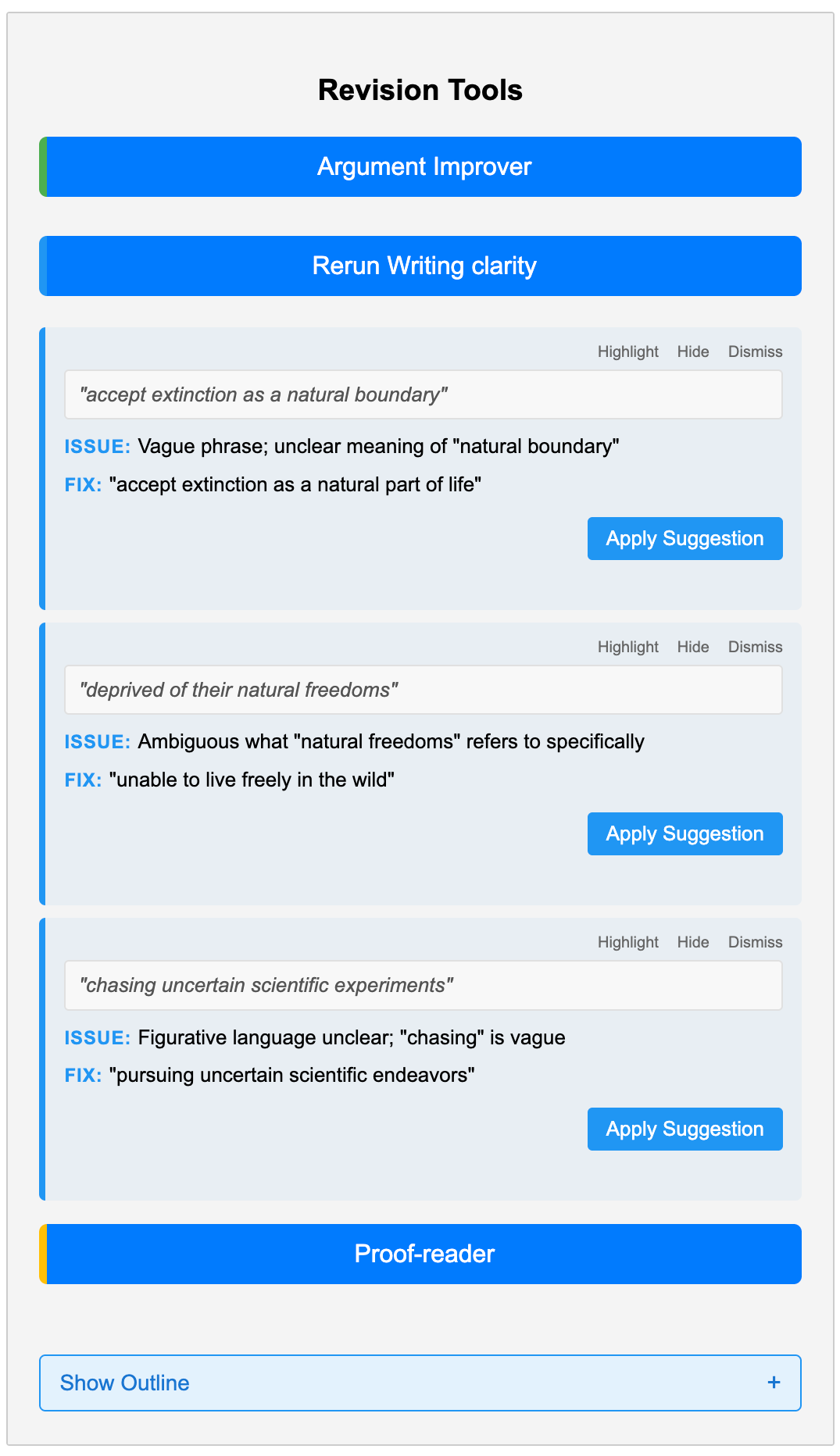}
        \caption{AI Revision Tool: Writing Clarity}
    \end{subfigure}
    \hfill
    \begin{subfigure}{0.32\textwidth}
        \includegraphics[width=\linewidth]{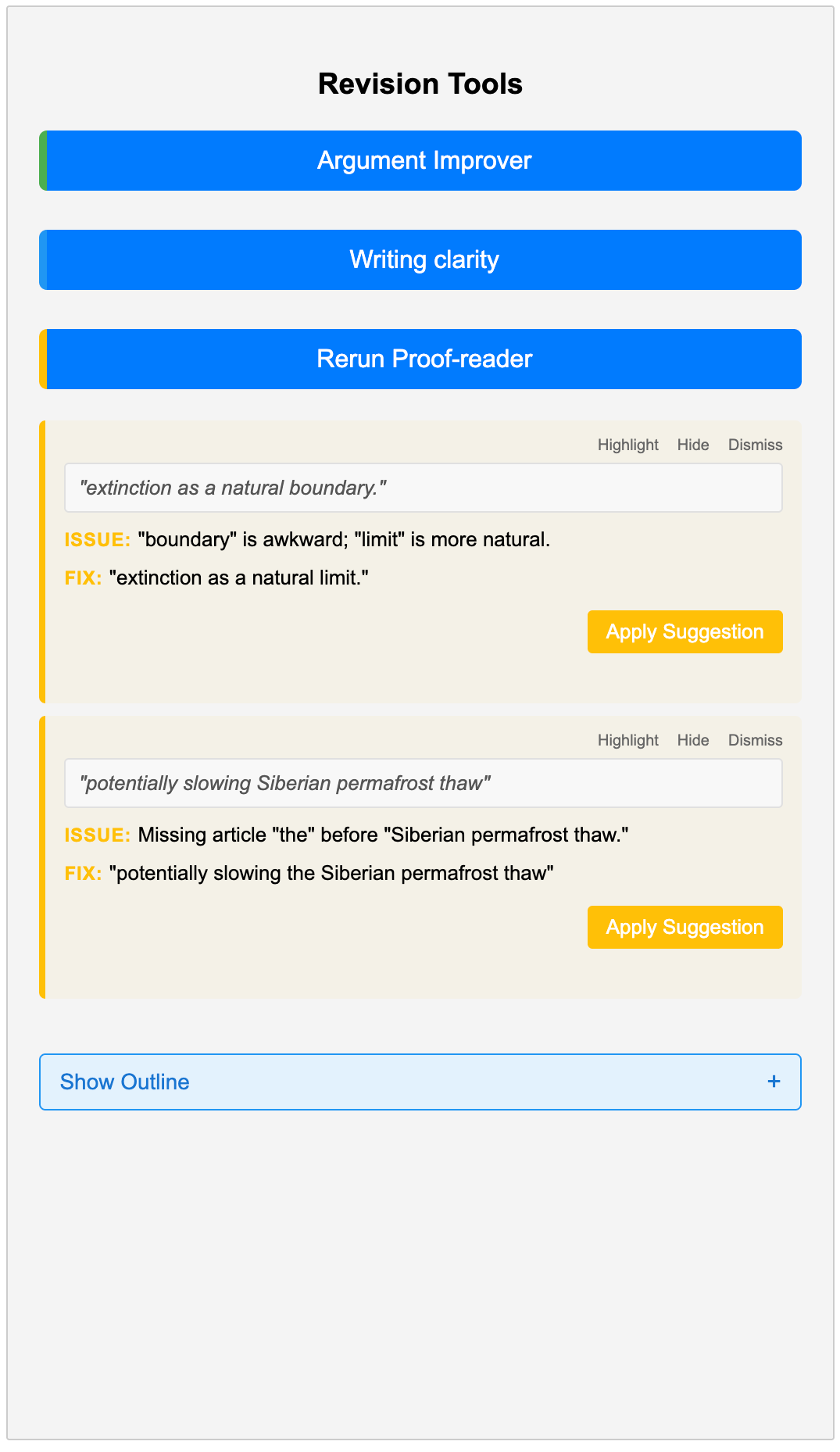}
        \caption{AI Revision Tool: Proof-reader}
    \end{subfigure}
    \caption{\textbf{Revision Stage (AI):} Examples of each of the three revision tools.}
\end{figure*}



\clearpage

\section{LLM Prompts}
\label{app:llmprompts}

\subsection{Planning Stage}

\subsubsection{AI Support for Planning Stage: Initial Request}

\begin{codebox}
\begin{myverbatim}
I am planning to write a 300 word argumentative essay on the topic: {promptText}.
Please give me 3 distinct thesis ideas for this essay:
- The first should argue in favor of the topic (pro position)
- The second should argue against the topic (con position)
- The third should be a nuanced or alternative perspective

Each idea must be extremely concise (10-15 words maximum) and make a clear argument.

Format your response with bullet points.

Do not include any additional explanation text before or after the bullet points.
Do not indicate which ideas are pro or con.
\end{myverbatim}
\end{codebox}

\subsubsection{AI Support for Planning Stage: Subsequent Requests}
\begin{codebox}
\begin{myverbatim}
Generate a single interesting thesis statement for a 300-word argumentative essay on the topic: {promptText}.
The statement must be extremely concise (10-15 words maximum) and make a clear argument.

Here are the existing ideas that have already been generated (DO NOT repeat these or anything similar):
{existingIdeas}

Provide ONLY the thesis statement with no additional text, introduction, or explanation.
\end{myverbatim}
\end{codebox}

\subsubsection{AI Support for Planning Stage: Full Outline Requests}
\begin{codebox}
\begin{myverbatim}
Create a very concise outline for a 300-word argumentative essay on this topic: {promptText}
The main thesis/idea is: {idea}

The outline must be extremely brief (no more than 20 lines total) with minimal formatting:
- One short introduction point
- 2-3 main points (no more than 15 words each)
- One short conclusion point

Use simple bullet points only. No sub-bullets or complex formatting.
\end{myverbatim}
\end{codebox}

\subsection{Drafting Stage}

\subsubsection{AI-generated Draft}
\begin{codebox}
\begin{myverbatim}
Please write a complete draft essay based on this outline and prompt.
      
Prompt: {promptText}
      
Outline:
{outline}
      
Please create a well-structured essay of approximately 300 words that follows this outline.
\end{myverbatim}
\end{codebox}

\clearpage

\subsection{Revision Stage}

\subsubsection{AI Revision Support: Proof-reader}
\begin{codebox}
\begin{myverbatim}
Please briefly review the following essay for the most critical typos, grammatical mistakes, and punctuation issues:
        
{draft}

Identify only 2-3 of the most important issues. For each issue:
1. Number the issue (Issue 1, Issue 2, etc.)
2. Quote the problematic phrase EXACTLY as it appears in the text (keep it under 10 words), using "double quotes".
3. Very briefly explain the problem in 5-10 words.
4. Suggest a concise fix. The Fix must be a direct replacement for the quoted text, with no instructions, options, or meta-comments. If the fix is to delete the text, return an empty string.

Format as:
### Issue 1
> "problematic text"
>
> **Issue**: Brief explanation
>
> **Fix**: "corrected version"

Be extremely concise. Only identify actual errors.
The quoted "problematic text" must appear EXACTLY as written in the original text - copy and paste the exact words, do not paraphrase or modify. This is critical for the tool to work properly.
If there are no issues, return "No issues found".
\end{myverbatim}
\end{codebox}

\subsubsection{AI Revision Support: Writing Clarity}
\begin{codebox}
\begin{myverbatim}
Please briefly review the following essay for the most unclear passages:

{draft}

Identify only 2-3 of the most problematic passages. For each issue:
1. Number the issue (Issue 1, Issue 2, etc.)
2. Quote the unclear phrase EXACTLY as it appears in the text (keep it under 10 words), using "double quotes".
3. Very briefly explain the clarity issue in 5-10 words.
4. Suggest a clearer alternative. The Fix must be a direct replacement for the quoted text, with no instructions, options, or meta-comments. If the fix is to delete the text, return an empty string.

Format as:
### Issue 1
> "unclear text"
>
> **Issue**: Brief explanation
>
> **Fix**: "clearer version"

Be extremely concise. Focus only on clarity issues.
The quoted "unclear text" must appear EXACTLY as written in the original text - copy and paste the exact words, do not paraphrase or modify. This is critical for the tool to work properly.
\end{myverbatim}
\end{codebox}

\clearpage

\subsubsection{AI Revision Support: Argument Improver}
\begin{codebox}
\begin{myverbatim}
Please briefly review the following essay for the most significant argument weaknesses:

{draft}

Identify only 2-3 of the most important argument issues. For each issue:
1. Number the issue (Issue 1, Issue 2, etc.)
2. Quote the relevant text that indicates approximately where the problem is (keep it under 15 words), using "double quotes". Quote a phrase EXACTLY as it appears in the text - copy and paste the exact words, do not paraphrase or modify.
3. Very briefly describe the argument weakness.
4. Provide a specific, actionable suggestion with concrete examples or specific steps.
5. If your suggestion involves adding new content, provide:
   - **Addition label**: a short (2-6 words) summary for a button label, e.g., 'add examples and case studies', 'add limitations', or 'add concrete examples'.
   - **Addition text**: the exact text to add (1-3 sentences).
   - **Insertion point**: a clear description of where to insert it. Use one of these formats:
     * 'After the sentence: "COMPLETE_SENTENCE"' - quote the ENTIRE sentence exactly as it appears in the text, from capital letter to period/question mark/exclamation mark
     * 'After paragraph X' - where X is the paragraph number (1, 2, 3, etc.)
   If your suggestion does not involve adding new content, leave these fields blank.

Format as:
### Issue 1
> "relevant text for improvement"
>
> **Issue**: Brief description of the argument weakness
>
> **Suggestion**: Specific, actionable advice with concrete examples or steps
>
> **Addition label**: add concrete examples
>
> **Addition text**: For instance, recent studies have shown that...
>
> **Insertion point**: After the sentence: "The argument lacks supporting evidence."

Be extremely concise. Focus only on argument quality and logic. For suggestions, provide specific examples, steps, or concrete details rather than general advice. The quoted text must appear EXACTLY as written in the original text - this is critical for the tool to work properly.

IMPORTANT: When using "After the sentence:" format, you MUST quote the COMPLETE sentence exactly as it appears in the text, from the first capital letter to the final punctuation mark. Do not truncate or abbreviate the sentence.
\end{myverbatim}
\end{codebox}

\clearpage
\section{Survey Questions}
\label{sec:surveyqs}

\def\CircleSLIDER#1#2{
\tikz[baseline=-0.1cm]{
 \coordinate (start) at (0,-0.1cm);
 \coordinate (end) at (#1,0.1cm);
 \coordinate (mark) at ($(start|-0,0)!#2!(end|-0,0)$);
 \fill[rounded corners=0.1cm, draw=gray, fill=lightgray] (start) rectangle (end);
 \fill[draw=gray, fill=blue!70] (mark) circle(.15) ;
}
}

\newcommand{\likertseven}{
\begin{enumerate}[label=\(\bigcirc\) \arabic*, leftmargin=1.3cm, align=left]
    \item \hspace{-10px}Strongly Disagree
    \item
    \item
    \item
    \item
    \item
    \item \hspace{-10px}Strongly Agree
\end{enumerate}
}

\subsection{Pre-Task Survey}

\paragraph{\textbf{Demographic questions:}}

\begin{itemize}
    \item What is your age? \textit{[18--25, 26--35, 36--45, 46--55, 56 or over]}
    \item What is your gender? \textit{[Male, Female, Non-binary / third gender, Prefer not to disclose]}
    \item What is your highest level of education? \textit{[High school diploma, Bachelor's degree, Master's degree, Doctorate, Other]}
    \item Is English your native language? \textit{[Yes, No]}
\end{itemize}





\paragraph{\textbf{Need for cognition questions:}}

[On a scale from 1 (Strongly Disagree) to 5 (Strongly Agree), rate the following:]

\begin{itemize}
    \item I prefer complex to simple problems.
    \item I like tasks that require little thought once I've learned them.
    \item Learning new ways to think doesn't excite me very much.
    \item I prefer a task that is intellectual and difficult to one that requires little thought.
\end{itemize}





\paragraph{\textbf{Technology acceptance questions:}}

[On a scale from 1 (Strongly Disagree) to 7 (Strongly Agree), rate the following:]

\begin{itemize}
    \item I find new technologies easy to use.
    \item Learning to use new technologies is easy for me.
    \item I find new technologies useful in my daily life.
    \item Using new technologies increases my productivity.
\end{itemize}





\paragraph{\textbf{AI tool usage questions:}}

\begin{itemize}
    \item How often do you use AI tools for writing (e.g., ChatGPT, Google Gemini, Claude)? \\ \textit{[Never, Less than once a month, About once a month, About once a week, Almost every day]}
    \item Which writing tasks have you used AI for? (Select all that apply) \\ \textit{[Brainstorming ideas, Drafting content, Revision / editing, I don't use AI for writing]}
\end{itemize}



\paragraph{\textbf{AI writing tools efficacy:}}

[On a scale from 1 (Strongly Disagree) to 7 (Strongly Agree), rate the following:]

\begin{itemize}
    \item I feel confident in my ability to use AI writing tools effectively.
    \item I believe AI writing tools can improve my writing quality.
\end{itemize}

\paragraph{\textbf{Writing efficacy:}}

[On a scale from 1 (Strongly Disagree) to 7 (Strongly Agree), rate the following:]

\begin{itemize}
    \item I can come up with creative ideas for my writing.
    \item I can organize my ideas in a logical way.
    \item I can write grammatically correct sentences.
    \item I can revise my own writing effectively.
    \item I can craft persuasive arguments in writing.
    \item I can write concisely without unnecessary details.
    \item I can structure an engaging introduction.
    \item Overall, I am confident in my writing abilities.
\end{itemize}

\subsection{Post-Task Survey}

\paragraph{\textbf{On a scale from 1 (Strongly Disagree) to 7 (Strongly Agree), rate the following:}}

\begin{itemize}
    \item I feel this piece of writing is truly mine.
    \item I feel this writing reflects my own voice and ideas.
    \item I feel I had complete control over the writing process.
    \item If I were to share this essay with a colleague, I would acknowledge AI support.
    \item I put a lot of effort into writing this essay.
    \item I feel that I actively chose all the arguments in this essay.
\end{itemize}

\paragraph{\textbf{Optional: Why did you (or did you not) feel this essay was truly yours?}}


\paragraph{\textbf{On a 0-100\% slider:}}

\begin{itemize}
    \item What percentage of the ideas in this text would you attribute to yourself versus AI?
    \item What percentage of the final text would you attribute to yourself versus AI?
\end{itemize}








\paragraph{\textbf{Optional: Why did you attribute that percentage of the text and ideas to yourself?}}

\paragraph{\textbf{Optional: Is there anything else you'd like to share about the process, interface, or your thoughts on authorship and ownership of the text?}}

\subsection{Psychometric Survey Results}
\label{app:psychresults}

Overall (pooled across conditions), participants reported the following for the psychometric surveys:

\begin{itemize}
    \item Need for Cognition (short 4-item form, 1-5 scale): \(M=2.97\), \(SD=0.41\), \(\alpha=.87\).
    \item Technology Acceptance (4 items, 1-7 scale): \(M=5.69\), \(SD=0.86\), \(\alpha=.85\).
    \item AI Writing Efficacy (2 items, 1-7 scale): \(M=5.38\), \(SD=1.30\), \(\alpha=.70\).
    \item Writing Self-Efficacy (8 items, 1-7 scale): \(M=5.63\), \(SD=0.92\), \(\alpha=.92\).
    
\end{itemize}

\section{Quality Control}
\label{app:qualitycontrol}

Given that AI usage for online experiments has been found to be quite high \cite{traylor2025ThreatAIChatbot}, we employed a number of methods to ensure participants completed our task without external AI support. We used a combination of keystroke logging, detection of paste events, manual inspection, and AI detection software. 
We used the Pangram AI detection API. See \url{https://www.pangram.com/} and the relevant technical report \cite{emi2024TechnicalReportPangram}. 

Note that it would be impossible to use exclusively AI detectors to determine external AI support as most participants were provided with AI support according to their condition. Additionally, AI detectors are known to have flaws \cite{elkhatat2023EvaluatingEfficacyAI}.

Instead, the participants were instructed that they had to write in the provided text box (as opposed to writing in a preferred text editor and pasting their results into the provided text box). If a participant typed fewer than 1000 characters throughout the course of the entire study or pasted more than 500 characters at once, then the participants' logs were manually inspected by one of the authors. (These thresholds were determined based on manual inspection of the first 50 participants.) Participants were excluded from the study if they pasted a large amount of text that did not come from within the browser (that is, was not a copy of their own text or provided AI support). Given that we explicitly instructed participants to type in the text box, participant submissions were rejected whenever there was too much pasted-in text, regardless of whether or not the text was determined to be AI generated. That said, the vast majority of large pastes appeared to be AI generated. 

Participants were also excluded if their submitted outline or draft was determined to be >99\% chance AI generated and they had unusual interaction patterns (e.g., very few keystrokes). (Very rarely, a submission came up as >99\% chance AI generated but the participant appeared to have typed it out, letter by letter, themselves, in a typical amount of time. In this case, we assumed the AI detector may be mistaken.)

18\% of submissions were excluded. An additional two participants were excluded due to data loss. 

\clearpage

\section{Automated Essay Grading}
\label{app:essay_grading}
 
\subsection{Grading Procedure}
 
A trained English-language teacher (the ``reference grader'') scored a
randomly selected subset of 50~essays on three dimensions adapted from
the IELTS Writing Task~2 Band Descriptors~\cite{IELTSBand}, each on a
0--9 integer scale:
\begin{enumerate}
  \item \textbf{Quality of argument} (mapped from IELTS ``Task Response'')~-- how well the writer develops and supports their ideas.
  \item \textbf{Coherence and cohesion}~-- how logically the essay flows and how ideas are connected.
  \item \textbf{Lexical resource and grammar} (combining the IELTS ``Lexical Resource'' and ``Grammatical Range and Accuracy'' descriptors)~-- vocabulary range, accuracy, and grammar use.
\end{enumerate}

Below are the IELTS band descriptors for bands 9, 6, and 3 -- note that the reference document used~\cite{IELTSBand} contains descriptors for all bands 0 to 9; we report three bands here to be representative and readers can reference the source document for complete descriptions. The trained English-language teacher was familiar with the band descriptors prior to the task, having graded actual IELTS essays previously (but was still provided the complete descriptors as reference).

\begin{table*}[h!]
\centering
\caption{IELTS Writing Task 2 Band Descriptors (Subset of Bands)}
\label{tab:ielts-bands}
\footnotesize
\begin{tabular}{|p{0.5cm}|p{4cm}|p{4cm}|p{4.5cm}|}
\hline
\textbf{Band} & \textbf{Task Response} & \textbf{Coherence \& Cohesion} & \textbf{Lexical Resource \& Grammar} \\
\hline
\textbf{9} &
Fully addresses all parts of the task; presents a fully developed position in aswer to the question with relevant, fully extended and well-supported ideas. &
Uses cohesion in such a way that it attracts no attention; skilfully manages paragraphing. &
Wide range of vocabulary with very natural and sophisticated control of lexical features; minor errors occur only as `slips'; full flexibility and precise use of grammar. \\
\hline
\textbf{6} &
Addresses all parts of the task although some parts may be more fully covered than others; presents a relevant position although the conclusions may become unclear or repetitive; presents relevant main ideas but some may be inadequately developed/unclear. &
Arranges information and ideas coherently and there is a clear overall progression; uses cohesive devices effectively, but cohesion within and/or between sentences may be faulty or mechanical; may not always use referencing clearly or appropriately; uses paragraphing, but not always logically. &
Uses an adequate range of vocabulary for the task; attempts to use less common vocabulary but with some inaccuracy; makes some errors in spelling and/or word formation, but they do not impede communication; uses a mix of simple and complex sentence forms; makes some errors in grammar and punctuation but they rarely reduce communication. \\
\hline
\textbf{3} &
Does not adequately address any part of the task; does not express a clear position; presents few ideas, which are largely undeveloped or irrelevant. &
Does not organise ideas logically; may use a very limited range of cohesive devices that may not indicate a logical relationship between ideas. &
Only a very limited range of words and expressions with very limited control of word formation and/or spelling; errors may severely distort the message. \\
\hline
\end{tabular}
\end{table*}
 
From this set of 50 human-graded essays, 30~were used as few-shot
calibration examples in the LLM grading prompt, and the remaining
20~were held out for validation.
 
\subsection{LLM Grading Model}
 
We used OpenAI's \texttt{o4-mini} reasoning model (with the
\texttt{reasoning.effort} parameter set to \texttt{high}).
This is a \emph{different} model from the \texttt{GPT-4o} instance that
provided AI writing support to participants during the study; the
grading model had no access to participants' interaction histories or
the experiment's AI-generated content.

\clearpage
 
\subsection{Grading Prompt}\label{app:grading-prompt}
 
The prompt template below was used to grade each essay.
The placeholder \texttt{\{rubric\_text\}} was replaced with the full
IELTS Task~2 band descriptors, and \texttt{\{examples\}} with the
30~human-graded essay--score pairs formatted as shown.

\begin{codebox}
\begin{myverbatim}
You are an expert IELTS examiner. Your task is to
evaluate student essays according to the IELTS Task 2
rubric.
 
You will give three separate scores (0-9):
1. Quality of argument - how well the student develops
   and supports their ideas.
2. Coherence and cohesion - how logically the essay
   flows and how ideas are connected.
3. Lexical resource and grammar - vocabulary range,
   accuracy, and grammar use.
 
Rubric (truncated to stay within context length):
{rubric_text}
 
Here are 30 graded examples for reference:
{examples}
 
Now grade the following essay: {essay_text}
 
Assign scores that reflect the standards from the
examples. Align with the reference grader.
 
1. Quality of argument: Look for weak reasoning,
   insufficient support, or underdeveloped ideas.
   Avoid giving "safe" midrange scores unless they
   truly match the examples.
2. Coherence and cohesion: Penalize logical gaps, poor
   transitions, or unclear organization.
3. Lexical resource and grammar: Evaluate normally
   based on the rubric.
 
Output format:
Quality of argument: X
Coherence and cohesion: Y
Lexical resource and grammar: Z

\end{myverbatim}
\end{codebox}

Each few-shot example was formatted as:
\begin{codebox}
\begin{myverbatim}
Essay:
{essay text}
 
Scores:
- Quality of argument: {score}
- Coherence and cohesion: {score}
- Lexical resource and grammar: {score}
---
\end{myverbatim}
\end{codebox}

\clearpage

\section{Thematic Analysis}
\label{app:thematic_analysis}


\begin{table}[h!]
\caption{Codebook from thematic analysis.}
\label{tab:thematic_analysis}
\small
\centering
\renewcommand{\arraystretch}{1.3}
\begin{tabular}[t]{lp{4cm}p{7cm}}
\toprule
\textbf{Affective Codes}\\
Code & Definition & Examples\\
\midrule
Emotional Resonance	&
Writing that resonates emotionally, aligns with values, or results in identification strengthens ownership.	& "I feel this essay is completely mine as it is something that I hold dear as am an advocate for conservation efforts. I also feel that all the ideas were originally mine" (\condition{none} condition) \\ 
&&"I wrote it from the bottom of my heart as this is something I am extremely passionate about." (\condition{plan} condition) \\ 
&& "I feel this essay is truly mine because it reflect my own thoughts and concerns about student debt. I wrote it in my own words using simple and honest ideas that I believe in." (\condition{none} condition)\\

Self-Efficacy& Belief in one’s own knowledge, competence, and background experience strengthens ownership.&"I feel I am very capable of discussing this topic since I have experienced it over the past few years. I understand both sides of the argument and wanted to let out how I feel about the perks of both sides." (\condition{none} condition) \\
Voice \& Expression &	Essays that match one's voice, style, or sense of personal expression strengthens ownership.	& "The outline was mine, but the voice, the writing style, was not. It can't truly be mine unless I was the one that created it all." (\condition{draft} condition)\\
&&
"Because I wrote it in my voice and it is my argument." (\condition{none} condition)\\
\midrule
\textbf{Behavioral Codes}\\
Effort \& Investment	& Putting in time, energy, or thought fosters pride and ownership.	&"I wrote all of it by myself, and am proud of the work I did." (\condition{none} condition)\\
&&
"I put effort into writing my personal views about genetic editing, but also got some inspirational from AI assistant to help me get started... I'm the one who form my own arguments and beliefs. " (\condition{plan} condition)\\

Integration of Ideas	& The way in which one makes integrates, expands, or changes ideas modifies feelings of ownership.	& \textbf{Participant didn't use AI ideas}: "The AI gave me a few ideas, but I ended up choosing to stick closely to my own thoughts" (\condition{plan} condition); "Although I got insight from AI but I never make use of it." (\condition{plan} condition)\\
&&
\textbf{AI repeated participant ideas}: "The AI did come up with outline ideas, but they matched/confirmed what I had already thought." (\condition{plan} condition)\\
&&
\textbf{AI contributed or changed ideas}: "About half of the points were mine and half of them were either introduced by the AI or I remember from the AI." (\condition{plan} condition); "The AI assistant added a number of points that I wasn't aware of or hadn't thought of." (\condition{draft} condition)\\
&&
\textbf{AI expanded on participant ideas}: "AI was able to provide some supplementary evidence to support my claim." (\condition{revision} condition); "Because the ideas are overall still in tact and fully mine... AI just expanded on the direction I was already going." (\condition{revision} condition)\\

\bottomrule
\end{tabular}
\end{table}

\clearpage

\section{Model Specifications and Planned Contrasts}
\label{app:model-details}

This appendix provides the full model specifications and contrast definitions for transparency and reproducibility; readers can follow the main results without the equations below.

\setcounter{equation}{0}
\renewcommand{\theequation}{\thesection.\arabic{equation}}

\subsection{Notation and prompt adjustment}
Across all research questions, let $i$ index participants, let $p_i$
denote the prompt variant, and let $c_i$ denote the assigned
condition:
\[
c_i \in \{
\textsc{no-ai},\;
\textsc{ai-plan},\;
\textsc{ai-draft},\;
\textsc{ai-revision}
\}.
\]
We use \textsc{no-ai} and a reference prompt $p_{\mathrm{ref}}$
(prompt a1) as baselines. Let $\mathcal{P}$ denote the set of
prompt variants.

When we report \emph{prompt-adjusted} (marginal) means by condition,
we average predictions equally across prompt variants:
\begin{equation}
\mu_c
=
\frac{1}{|\mathcal{P}|}
\sum_{p \in \mathcal{P}}
\mathbb{E}\!\left[y_i \mid c_i = c,\; p_i = p\right].
\label{eq:prompt-adjusted-mean}
\end{equation}
This equal weighting ensures that stage comparisons are not driven by
small differences in prompt mix across conditions.

\subsection{RQ1: Ownership}
Self-reported primary ownership is treated as approximately continuous
on a 1--7 scale. We estimate a linear model with prompt fixed effects
and heteroskedasticity-robust standard errors:
\begin{equation}
\begin{aligned}
y_i
&=
\alpha
+ \beta_{\text{plan}}\,\mathbf{1}\{c_i=\textsc{ai-plan}\}
+ \beta_{\text{draft}}\,\mathbf{1}\{c_i=\textsc{ai-draft}\} \\
&\quad
+ \beta_{\text{rev}}\,\mathbf{1}\{c_i=\textsc{ai-revision}\}
+ \sum_{p \in \mathcal{P}\setminus\{p_{\mathrm{ref}}\}}
\gamma_p\,\mathbf{1}\{p_i=p\}
+ \varepsilon_i,
\label{eq:own-explicit}
\end{aligned}
\end{equation}
where $y_i$ is participant $i$'s ownership rating and
$\varepsilon_i$ is a mean-zero error term. For exposition we write
$\varepsilon_i \sim \mathcal{N}(0,\sigma^2)$, but all inference uses
robust standard errors, so neither normality nor homoskedasticity
is assumed.

Here, $\alpha$ is the mean ownership in the \textsc{no-ai} condition
at the reference prompt; $\beta_{\text{plan}}$, $\beta_{\text{draft}}$,
and $\beta_{\text{rev}}$ are condition differences relative to
\textsc{no-ai}; and $\gamma_p$ captures the mean shift for prompt $p$
relative to $p_{\mathrm{ref}}$.

\paragraph{Planned contrasts (C1--C3).}
We evaluate three pre-specified centered contrasts on the
prompt-adjusted means $\mu_c$:
\begin{align}
\mathrm{C1}: \quad
&\mu_{\textsc{no-ai}}
- \tfrac{1}{3}\!\left(
\mu_{\textsc{ai-plan}} +
\mu_{\textsc{ai-draft}} +
\mu_{\textsc{ai-revision}}
\right), \nonumber\\
\mathrm{C2}: \quad
&\mu_{\textsc{ai-draft}}
- \tfrac{1}{2}\!\left(
\mu_{\textsc{ai-plan}} +
\mu_{\textsc{ai-revision}}
\right), \nonumber\\
\mathrm{C3}: \quad
&\mu_{\textsc{ai-plan}} - \mu_{\textsc{ai-revision}}.
\label{eq:planned-contrasts}
\end{align}

The corresponding contrast weight vectors are
\begin{equation}
\begin{aligned}
&\text{over }
(\mu_{\textsc{no-ai}}, \mu_{\textsc{ai-plan}},\\
&\qquad \mu_{\textsc{ai-draft}}, \mu_{\textsc{ai-revision}})
\text{ are:}
\end{aligned}
\end{equation} are
\begin{align}
\mathbf{w}_{\text{C1}}
&=
\left(
1,\; -\tfrac{1}{3},\; -\tfrac{1}{3},\; -\tfrac{1}{3}
\right), \nonumber\\
\mathbf{w}_{\text{C2}}
&=
\left(
0,\; -\tfrac{1}{2},\; 1,\; -\tfrac{1}{2}
\right), \nonumber\\
\mathbf{w}_{\text{C3}}
&=
\left(0,\; 1,\; 0,\; -1\right).
\label{eq:contrast-weights}
\end{align}

\subsection{RQ2: Attribution to AI (ideas vs.\ text)}
For each attribution dimension $a$
(\textit{ideas\_AI} and \textit{text\_AI}),
we model the self-reported percentage attributed to AI as a
proportion in $[0,1]$ and fit a generalized linear model with
a logit link:
\begin{equation}
\begin{aligned}
\operatorname{logit}\!\left(\mu_i^{(a)}\right)
&=
\alpha^{(a)}
+ \beta_{\text{plan}}^{(a)}\,\mathbf{1}\{c_i=\textsc{ai-plan}\} \\
&\quad
+ \beta_{\text{draft}}^{(a)}\,\mathbf{1}\{c_i=\textsc{ai-draft}\}
+ \beta_{\text{rev}}^{(a)}\,\mathbf{1}\{c_i=\textsc{ai-revision}\} \\
&\quad
+ \sum_{p \in \mathcal{P}\setminus\{p_{\mathrm{ref}}\}}
\gamma_p^{(a)}\,\mathbf{1}\{p_i=p\}.
\label{eq:attr-explicit}
\end{aligned}
\end{equation}
Here,
$\mu_i^{(a)}=\mathbb{E}[y_i^{(a)} \mid c_i,p_i]$
is the conditional mean proportion attributed to AI on dimension $a$,
and
$\operatorname{logit}(x)=\log\!\bigl(\tfrac{x}{1-x}\bigr)$.
The intercept $\alpha^{(a)}$ is the baseline for \textsc{no-ai} at the
reference prompt, the $\beta^{(a)}$ terms are condition effects
relative to \textsc{no-ai}, and the $\gamma_p^{(a)}$ terms are prompt
fixed effects.

Because these outcomes are self-reported percentages rather than
literal binomial counts, we use a quasi-binomial variance and robust
standard errors. For reporting, we back-transform fitted values to
the response scale and present prompt-adjusted marginal means for
each stage, with 95\% confidence intervals. We then apply the same
three planned contrasts (C1--C3) separately to \emph{ideas} and
\emph{text}, expressing differences in percentage points.

\subsection{RQ3: Essay Quality}
We model essay quality as a continuous outcome on a 0--9 scale,
using the same stage factor and prompt fixed effects as in RQ1,
plus two covariates capturing task effort: final word count and total
time on task. Let $q_i$ denote the composite quality score for
participant $i$. We estimate
\begin{equation}
\begin{aligned}
q_i
&=
\alpha
+ \beta_{\text{plan}}\,\mathbf{1}\{c_i=\textsc{ai-plan}\}
+ \beta_{\text{draft}}\,\mathbf{1}\{c_i=\textsc{ai-draft}\} \\
&\quad
+ \beta_{\text{rev}}\,\mathbf{1}\{c_i=\textsc{ai-revision}\}
+ \sum_{p \in \mathcal{P}\setminus\{p_{\mathrm{ref}}\}}
\gamma_p\,\mathbf{1}\{p_i=p\} \\
&\quad
+ \delta_1\,\mathrm{WC}_i
+ \delta_2\,\mathrm{Time}_i
+ \varepsilon_i.
\label{eq:qual-explicit}
\end{aligned}
\end{equation}
Here, $\alpha$ is the intercept,
$\beta_{\text{plan}}$, $\beta_{\text{draft}}$, and $\beta_{\text{rev}}$
are stage effects relative to \textsc{no-ai},
$\gamma_p$ are prompt fixed effects,
$\mathrm{WC}_i$ is final word count,
and $\mathrm{Time}_i$ is total minutes on task.
The error term $\varepsilon_i$ has mean zero, and inference again
uses heteroskedasticity-robust standard errors.

For reporting, we present adjusted marginal means by stage,
computed by averaging equally over prompts and holding the numeric
covariates at their sample means. We then evaluate the same three
planned contrasts (C1--C3) directly on the 0--9 quality scale.
\subsection{Estimation and reporting}
All models include prompt fixed-effects ($\gamma_{p}$) to absorb the effects of varying essay topic. For linear models we use OLS with heteroskedasticity-consistent standard errors; for proportional outcomes we use a quasi-binomial GLM with a logit link. Condition-level marginal means and planned contrasts are computed using estimated marginal means (on the response scale where applicable). We report two-sided 95\% confidence intervals and $p$-values for the small set of prespecified contrasts, which are reused consistently across RQ1--RQ3.

Analyses were conducted in \textsf{R}  using \textit{sandwich}  and \textit{emmeans}  for robust inference, marginal means, and planned contrasts.

\clearpage
\section{Results}
\label{app:results}

\begin{figure*}[ht]
    \centering

    \begin{subfigure}{\textwidth}
        \centering
        \includegraphics[width=0.8\linewidth]{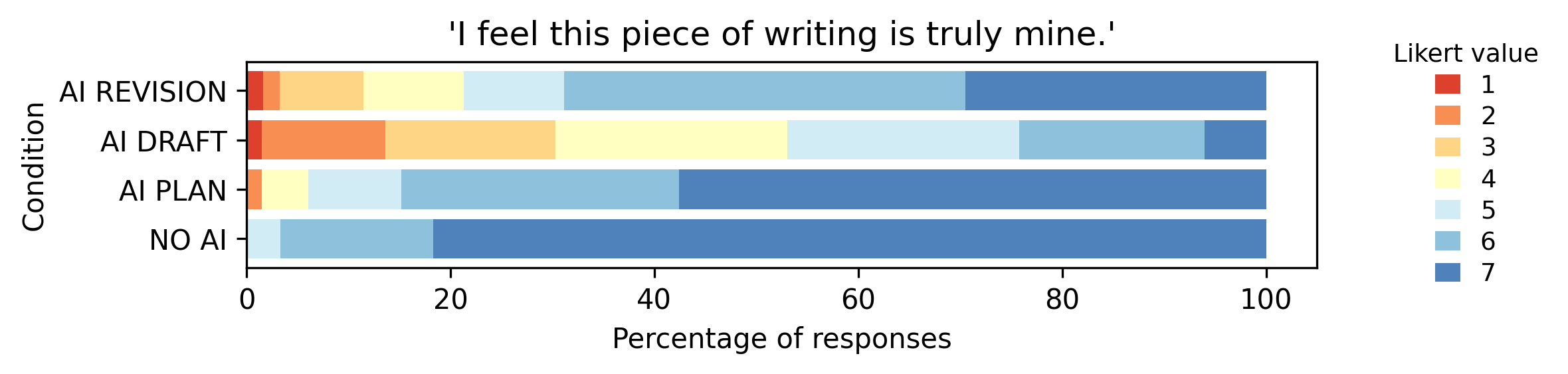}
        \caption{Likert scale responses by condition}
        \label{fig:likert}
    \end{subfigure}

    \vspace{1em}

    \begin{subfigure}{0.47\textwidth}
        \centering
        \includegraphics[height=5cm]{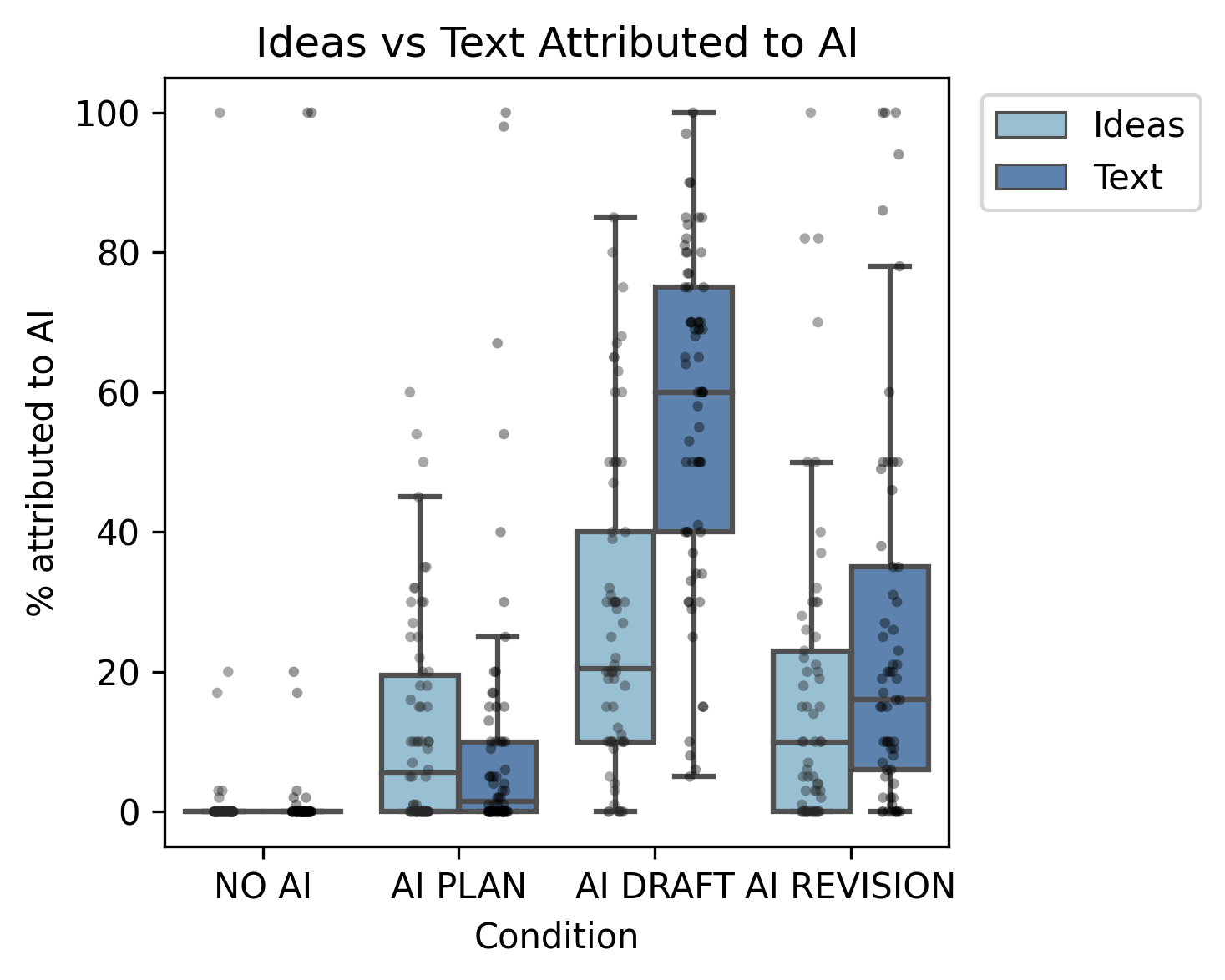}
        \caption{Ideas vs Text attributed to AI}
        \label{fig:ideasvtext}
    \end{subfigure}
    \hspace{0.04\textwidth} 
    \begin{subfigure}{0.35\textwidth}
        \centering
        \includegraphics[height=5cm]{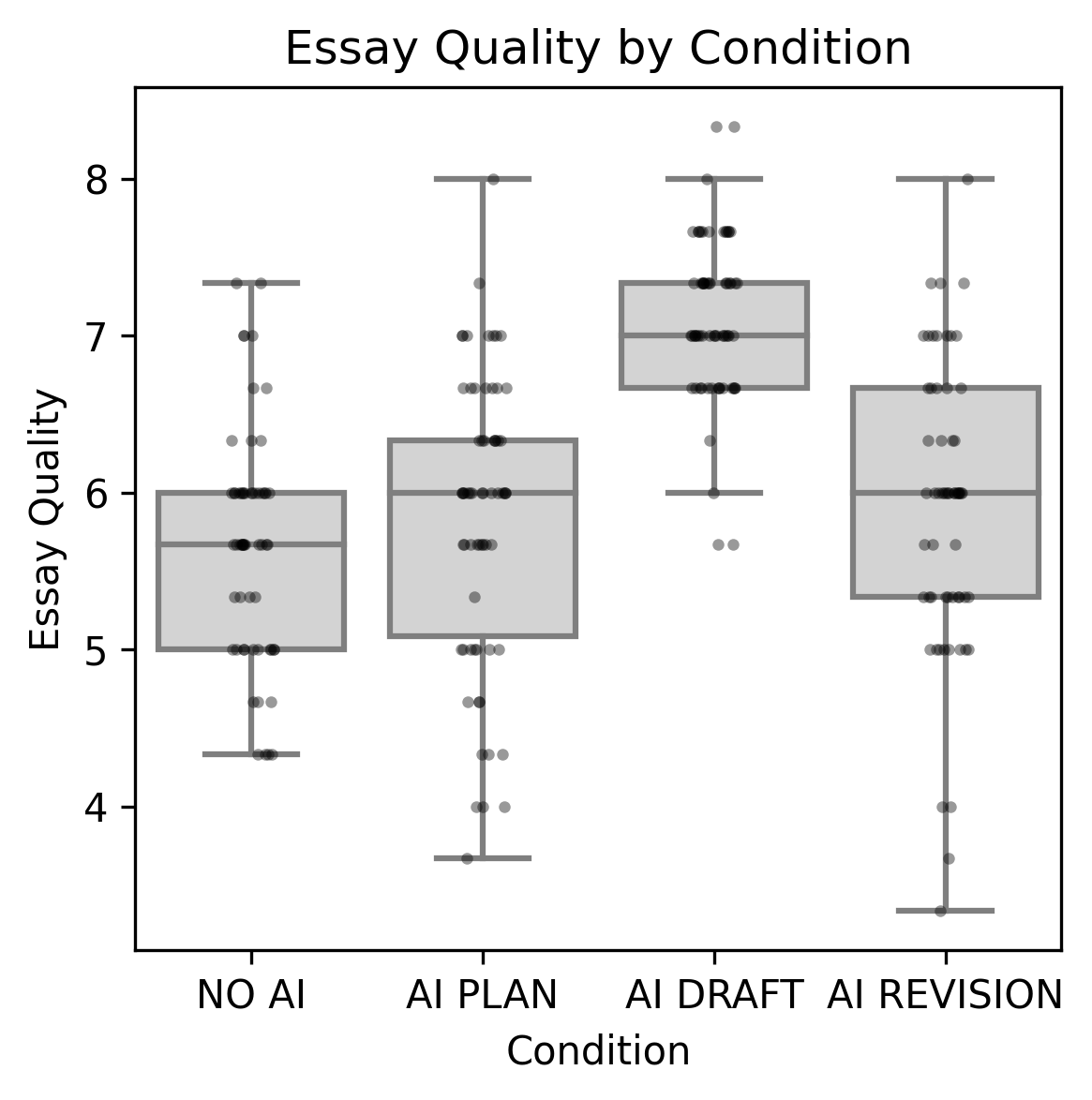}
        \caption{Essay quality scores by condition}
        \label{fig:essayquality}
    \end{subfigure}

    \caption{Survey and essay results across conditions. 
        (a) Likert scale distributions, 
        (b) self-reported attribution of ideas vs text, and 
        (c) average essay quality scores.}
    \label{fig:results-descriptors}
\end{figure*}

\autoref{tab:consolidated_means_by_rq} shows all adjusted means and confidence intervals for the three research questions.
\autoref{fig:results-descriptors} shows the data distributions for all our main results, split by condition.

\begin{table}[htbp]
  \centering
  \small
  \caption{Adjusted means with 95\% CIs by stage for the three research questions. RQ1 Ownership is on the original 1–7 Likert scale; RQ2 shows the percent attributed to AI for \emph{ideas} and \emph{text}; RQ3 Quality is on a 0–9 rubric scale. Note: ``Adjusted mean'' = stage average after holding prompt differences constant (averaging over prompt variants). RQ3 additionally adjusts for final word count and total minutes. Confidence intervals use robust standard errors.}
  \label{tab:consolidated_means_by_rq}

  \vspace{0.5em}

  \begin{subtable}{.35\textwidth}
    \centering
    \caption{RQ1 — Ownership (Likert, 1–7)}
    \label{tab:rq1_own_means}
    \begin{tabular}{lcc}
      \toprule
      \textbf{Stage} & \textbf{Adjusted Mean} & \textbf{95\% CI} \\
      \midrule
      \textsc{no-ai}        & 6.74 & [6.61, 6.88] \\
      \textsc{ai–plan}      & 6.30 & [6.04, 6.56] \\
      \textsc{ai–draft}     & 4.29 & [3.92, 4.66] \\
      \textsc{ai–revision}  & 5.57 & [5.19, 5.94] \\
      \bottomrule
    \end{tabular}
  \end{subtable}
  
  \vspace{0.75em}
  
  \begin{subtable}{.48\textwidth}
    \centering
    \caption{RQ2 — Attribution to AI (\%)}
    \label{tab:rq2_attr_means}
    \begin{tabular}{lcccc}
      \toprule
      \multirow{2}{*}{\textbf{Stage}} & \multicolumn{2}{c}{\textbf{Ideas to AI}} & \multicolumn{2}{c}{\textbf{Text to AI}} \\
      \cmidrule(lr){2-3}\cmidrule(lr){4-5}
       & \textbf{Adj. Mean} & \textbf{95\% CI} & \textbf{Adj. Mean} & \textbf{95\% CI} \\
      \midrule
     \textsc{no-ai}      & 2.4  & [0.9, 6.3]   & 4.0  & [1.7, 9.0] \\
      \textsc{ai–plan}    & 11.4 & [7.4, 17.3]  & 9.6  & [5.7, 15.6] \\
      \textsc{ai–draft}    & 26.9 & [20.6, 34.4] & 56.9 & [48.4, 65.0] \\
      \textsc{ai–revision} & 16.1 & [11.0, 22.9] & 24.3 & [17.6, 32.5] \\
      \bottomrule
    \end{tabular}
  \end{subtable}

  \vspace{0.75em}

  \begin{subtable}{\textwidth}
    \centering
    \caption{RQ3 — Essay Quality (rubric, 0–9)}
    \label{tab:rq3_quality_means}
    \begin{tabular}{lcc}
      \toprule
      \textbf{Stage} & \textbf{Adjusted Mean} & \textbf{95\% CI} \\
      \midrule
      \textsc{no-ai}       & 5.69 & [5.50, 5.88] \\
      \textsc{ai–plan}    & 5.85 & [5.62, 6.08] \\
     \textsc{ai–draft}      & 7.27 & [7.06, 7.48] \\
     \textsc{ai–revision}  & 5.79 & [5.55, 6.03] \\
      \bottomrule
    \end{tabular}
  \end{subtable}

  \vspace{0.25em}
  \footnotesize
  
\end{table}


\clearpage

\subsection{AI Support Usage}

\autoref{tab:engagement_by_stage} shows more detail on how people engaged with the AI support in the \condition{plan} and \condition{revision} conditions.

\begin{table*}[h!]
  \centering
  \small
  \caption{Engagement with AI assistance. We report metrics from the \textsc{ai-plan} and \textsc{ai-revision} conditions as these had interactive elements; the \textsc{ai-draft} condition required all participants to generate an AI draft, so there are no condition-specific engagement metrics. In \textsc{ai-plan}, 43 of 66 participants (65.2\%) requested at least one AI idea; in \textsc{ai-revision}, 51 of 61 participants (83.6\%) ran at least one revision tool.}
  \label{tab:engagement_by_stage}

  \begin{subtable}{.47\textwidth}
    \centering
    \begin{tabular}{lc}
      \toprule
      \textbf{Metric} &  \textbf{Mean (SD)} \\
      \midrule
      Outline ideas requested (binary)           &  0.65 (-)\\
      Outline detail shown (count)               & 0.79 (0.85) \\
      Outline ideas pasted (count)               & 0.41 (0.93) \\
      Outline characters inserted (chars)        & 101 (231) \\
      Outline character adoption share (\%)           & 5.7 (14.0) \\
      \bottomrule
    \end{tabular}
    \caption{Planning condition (\condition{plan}). Participants could request single sentence ideas and request, for an already generated idea, a detailed outline. \textit{Outline char adoption share} reports the percent of final characters in the submitted outline that came from AI suggestions.}

    \label{tab:engage_plan}
  \end{subtable}%
    \hspace{0.03\textwidth} 
  \begin{subtable}{.47\textwidth}
    \centering
    \begin{tabular}{lc}
      \toprule
      \textbf{Metric} &  \textbf{Mean (SD)} \\
      \midrule
      Revision tools run (binary)                & 0.84 (-) \\
      Revision suggestions shown (count)         & 13.90 (11.80) \\
      Revision suggestions applied (count)       & 7.36 (6.74) \\
      Revision acceptance (applied / shown) (\%) & 52.7 (25.8) \\
      Revision characters inserted (chars)       & 658 (562) \\
      Revision character share (\%)              & 26.5 (21.0) \\
      \bottomrule
    \end{tabular}
    \caption{Revision condition (\condition{revision}). Participants could request suggestions from three tools (argument improvements, writing clarity, proof reader) as they liked. \textit{Revision character share} reports the percent of characters in the submitted essay that came from suggestions.}
    \label{tab:engage_rev}
  \end{subtable}

\end{table*}

\clearpage

\subsection{Robustness to Tool Engagement}
\label{app:robustness-engagement}

Because AI tool use was optional in \condition{plan} and \condition{revision}, the primary analyses in Section~5 compare participants by assigned condition regardless of engagement. Table~\ref{tab:robustness-engagement} repeats the main adjusted means after restricting to participants who actively used the available tools.

\paragraph{Definitions.} We define active tool users as participants who requested at least one AI idea (\condition{plan}) or ran at least one revision tool (\condition{revision}). This reduces \condition{plan} from $n=66$ to $n=43$ (65\%) and \condition{revision} from $n=61$ to $n=51$ (84\%). \condition{no-ai} and \condition{draft} are unchanged.

\begin{table}[h]
\centering
\caption{Adjusted means restricted to active tool users. Compare to Table~\ref{tab:consolidated_means_by_rq} (full sample).}
\label{tab:robustness-engagement}
\small
\begin{tabular}{lcccc}
\toprule
& \textsc{no-ai} & \textsc{ai-plan} & \textsc{ai-draft} & \textsc{ai-revision} \\
& $(n=60)$ & $(n=43)$ & $(n=66)$ & $(n=51)$ \\
\midrule
\multicolumn{5}{l}{\textit{RQ1: Ownership (1--7 scale)}} \\
\quad Adjusted Mean & 6.73 & 6.16 & 4.28 & 5.40 \\
\quad 95\% CI & [6.59, 6.87] & [5.84, 6.47] & [3.92, 4.65] & [4.99, 5.81] \\
\midrule
\multicolumn{5}{l}{\textit{RQ2: Ideas attributed to AI (\%)}} \\
\quad Adjusted Mean & 2.4 & 16.2 & 26.9 & 17.1 \\
\quad 95\% CI & [0.0, 4.8] & [9.1, 23.4] & [19.8, 34.0] & [10.3, 23.9] \\
\midrule
\multicolumn{5}{l}{\textit{RQ2: Text attributed to AI (\%)}} \\
\quad Adjusted Mean & 4.0 & 12.3 & 57.0 & 23.7 \\
\quad 95\% CI & [0.8, 7.1] & [5.9, 18.6] & [49.0, 65.0] & [16.0, 31.4] \\
\midrule
\multicolumn{5}{l}{\textit{RQ3: Essay Quality (0--9 scale)}} \\
\quad Adjusted Mean & 5.67 & 5.93 & 7.25 & 5.88 \\
\quad 95\% CI & [5.48, 5.87] & [5.63, 6.23] & [7.03, 7.46] & [5.60, 6.15] \\
\bottomrule
\end{tabular}
\end{table}

Overall, this restriction does not change the substantive conclusions. Ownership retains the same stage ordering (highest in \textsc{no-ai}, highest among AI stages in \condition{plan}, lowest in \condition{draft}). The planned contrasts C1 and C2 retain their direction and remain statistically reliable across outcomes; C3 remains statistically reliable for ownership and text attribution, but not for ideas attribution or essay quality, mirroring the full-sample pattern. For ownership specifically, C3 (\condition{plan} vs.\ \condition{revision}) is slightly larger among active users ($+0.76$, $p=.003$) than in the full sample ($+0.73$, $p=.0016$).


\end{document}